\documentclass{aastex}
\usepackage{rotating}
\usepackage{emulateapj5}

\slugcomment{To appear in the Astrophysical Journal, Aug 1, 2012}

\shorttitle{Eclipse of WASP-33b}
\shortauthors{Deming et al.}

\begin{document}


\title{Infrared Eclipses of the Strongly Irradiated Planet WASP-33b, and Oscillations of its Host Star}


\author{Drake~Deming\altaffilmark{1,2}, Jonathan~D.~Fraine\altaffilmark{1,2}, 
Pedro~V.~Sada\altaffilmark{3,2}, Nikku~Madhusudhan\altaffilmark{4,5}, Heather~A.~Knutson\altaffilmark{6}, 
Joseph~Harrington\altaffilmark{7}, Jasmina~Blecic\altaffilmark{7}, Sarah~Nymeyer\altaffilmark{7,8},
Alexis~M.~S.~Smith\altaffilmark{9}, \& Brian~Jackson\altaffilmark{10}}


\altaffiltext{1}{Department of Astronomy, University of Maryland, College Park, MD 20742; ddeming@astro.umd.edu}
\altaffiltext{2}{Visiting Astronomer, Kitt Peak National Observatory, National Optical Astronomy 
  Observatory, which is operated by the Association of Universities for Research in Astronomy under 
  cooperative agreement with the National Science Foundation}
\altaffiltext{3}{Universidad de Monterrey, Monterrey, M\'exico}
\altaffiltext{4}{Department of Astrophysical Sciences, Princeton University, Princeton, NJ 08544-1001}
\altaffiltext{5} {Present address: Yale Center for Astronomy \& Astrophysics, Yale University, New Haven, CT 06511}
\altaffiltext{6} {Division of Geological and Planetary Sciences, California Institute of Technology, Pasadena, CA 91125}
\altaffiltext{7} {Planetary Sciences Group, Department of Physics, University of Central Florida, Orlando, FL 32816-2385}
\altaffiltext{8} {Present address: Department of Earth and Space Sciences, University of California at Los Angeles, 
                   Los Angeles, CA 90095-1567}
\altaffiltext{9} {Astrophysics Group, Keele University, Staffordshire ST5 5BG, UK}
\altaffiltext{10} {Department of Terrestrial Magnetism, Carnegie Institution of Washington, Washington, DC 20015}


\begin{abstract}
We observe two secondary eclipses of the strongly irradiated
transiting planet WASP-33b, in the $K_s$ band at 2.15\,$\mu$m, and one
secondary eclipse each at 3.6\,$\mu$m and 4.5\,$\mu$m using Warm
Spitzer.  This planet orbits an A5V $\delta$-Scuti star that is known
to exhibit low amplitude non-radial p-mode oscillations at about 0.1\%
semi-amplitude.  We detect stellar oscillations in all of our infrared
eclipse data, and also in one night of observations at J-band
(1.25\,$\mu$m) out of eclipse.  The oscillation amplitude, in all
infrared bands except $K_s$, is about the same as in the optical.
However, the stellar oscillations in $K_s$ band (2.15\,$\mu$m) have
about twice the amplitude (0.2\%) as seen in the optical, possibly
because the Brackett-$\gamma$ line falls in this bandpass. As regards
the exoplanetary eclipse, we use our best-fit values for the eclipse
depth, as well as the 0.9\,$\mu$m eclipse observed by \citet{smith},
to explore possible states of the exoplanetary atmosphere, based on
the method of \citet{madhu09}. On this basis we find two possible
states for the atmospheric structure of WASP-33b. One possibility is a
non-inverted temperature structure in spite of the strong irradiance,
but this model requires an enhanced carbon abundance ($C/O>1$).  The
alternative model has solar composition, but an inverted temperature
structure. Spectroscopy of the planet at secondary eclipse, using a
spectral resolution that can resolve the water vapor band structure,
should be able to break the degeneracy between these very different
possible states of the exoplanetary atmosphere.  However, both of
those model atmospheres absorb nearly all of the stellar irradiance
with minimal longitudinal re-distribution of energy, strengthening the
hypothesis of \citet{cowan} that the most strongly irradiated planets
circulate energy poorly.  Our measurement of the central phase of the
eclipse yields $e\,cos\,\omega=0.0003\pm0.00013$, which we regard as
being consistent with a circular orbit.
\end{abstract}

%% for the journal to which y ou are submitting your paper to determine


\section{Introduction}

Extrasolar planets that orbit close to their stars are subject to an
intense flux of stellar irradiation.  The rotation of a very close-in
planet is expected to become tidally locked to its orbital period on
an astrophysically short time scale \citep{guillot}.  Consequently,
the most close-in exoplanets will receive stellar irradiation
exclusively on their star-facing hemispheres.  The resulting heating
is believed to be distributed by strong zonal winds \citep{showman},
but the dynamics of the zonal re-distribution, and therefore the
overall energy budget of the planet, are affected by the vertical
temperature structure of the planetary atmosphere.  Many close-in
planets exhibit inverted temperature structures \citep{knutson08,
seager-deming}, probably driven by radiative absorption in a high
altitude layer of the atmosphere \citep{burrows07}.  The nature of the
absorber has been actively discussed \citep{fortney, spiegel}, but
remains unknown.  

One promising avenue of investigation is to look for correlations
between the planetary temperature inversion and the stellar flux at
ultraviolet (UV) wavelengths \citep{knutson10}.  Stellar UV radiation
has the potential to dissociate absorbing molecular species,
and to create (or destroy) absorbers via photochemistry \citep{zahnle}.  
The spectral distribution of UV flux may be critical to the inversion
phenomenon.  Therefore it is desirable to investigate planets orbiting
strong sources of far-UV radiation (i.e., magnetically active stars),
as well as planets receiving irradiation by thermal UV radiation
(i.e., hot stars).

An important planet in the latter category is WASP-33b, that orbits an
A-type $\delta$-Scuti star with an orbital period of 1.22 days
\citep{collier-cameron, herrero}.  The large radius and high
temperature of an A-type star produce stronger irradiance than would
be the case for a solar-type star.  Although only an upper limit is
available for the mass of WASP-33b \citep{collier-cameron}, the planet
is important because it is among the most strongly irradiated planets.
In contrast to other strongly irradiated planets such as WASP-12,
\citep{cowan12, crossfield, zhao, campo, croll, madhu11a}, little is
currently known about the response of WASP-33b to the strong stellar
irradiation.  \citet{smith} measured the thermal emission of WASP-33b
from secondary eclipse observations at $0.9$\,$\mu$m, but there are
currently no reported detections of the planet at wavelengths longward
of 1\,$\mu$m.

Stellar intensity oscillations of WASP-33 are seen with 0.1\%
semi-amplitude at optical wavelengths \citep{herrero}.  The stellar
oscillations may exhibit a greater or lesser amplitude at infrared
(IR) wavelengths.  The dependence of the oscillation amplitude on
wavelength potentially carries information on the physics of the
oscillations in the stellar atmosphere.

In this paper, we report measurement of the thermal emission from
WASP-33b, based on ground-based observations of two secondary eclipses
in the $K_s$ band ($2.15$\,$\mu$m), space-borne observations of
eclipses at 3.6- and 4.5\,$\mu$m by Warm Spitzer, and measurement of
the intensity oscillations of the star in all these IR bands, as well
as in J-band ($1.25$\,$\mu$m).  We describe the observations and
extraction of photometry from the data in Sec.~2~\&~3.  In Sec.~4, we
analyze the data to determine the parameters of the planet's eclipse
and the oscillatory properties of the star.  We explore and discuss
the implications of our results in Sec.~5, and Sec.~6 summarizes our
results.

\section{Observations}

\subsection{Ground-based Observations}

We observed secondary eclipses of WASP-33b on UT 10 and 16 October,
2011, using the FLAMINGOS infrared HgCdTe imager at the Cassegrain
focus of the 2.1-meter telescope at Kitt Peak National
Observatory. The sky on both nights was cloudless, with excellent
photometric conditions prevailing, especially on 10 October.  The
observations used a $K_s$ filter, and we defocused the telescope so
that the diameter of stellar images - measured at the half-intensity
level - was 30-pixels (18 arc-sec). This substantial defocus improves
the photometric precison and photon-collection efficiency for this
bright star (WASP-33 has V=8.3). We compensated for drift in the
telescope defocus using manual updates at approximately 30
minute intervals, based on a known formula that relates focus position to
temperature and zenith distance.  In that way, we attempted to
maintain the greatest possible image stability.

The $20 \times 20$ arc-min field of view of FLAMINGOS ($2048 \times
2048$ pixels) provided 5 comparison stars imaged simultaneously with
WASP-33.  We obtained a nearly continuous sequence of 20-second
exposures on each night, amounting to 725 exposures on 10 October and
574 exposures on 16 October.  Including the overhead of reading the
detector and writing FITS files, the observational cadence was 45
seconds per exposure.  We verified that the times written to the FITS
headers were free of clock errors (to about 1 second precision) by
comparing the header values to manual timing made using a
web-displayed UT clock.  In addition to WASP-33, we observed multiple
sky exposures with position offsets, that were used to construct a
flat-field frame by median-combining the offset sky exposures.  We
also observed dark frames using the same exposure times as for the
stellar and sky-flat observations.

All of our WASP-33 observations used a quad-detector off-axis guide camera,
sensitive to optical radiation, and producing real-time pointing
corrections for the telescope.  The guider was very effective at
damping image motion on time scales of minutes, but differential
refraction between the optical and the infrared leads to a slow drift
in stellar positions, amounting to about 1 arc-sec over 60-degrees of
zenith distance.  This slow positional drift has only a small effect
on our photometry.

During the observations, we noted a significant instability in the
infrared signal from the FLAMINGOS detector, that occurred in response
to the changing position of the telescope.  In order to diagnose and
correct for these instabilites, we obtained a sequence of K-band
exposures during the day, with the dome closed and dark, and the
primary mirror cover closed.  These 2-second exposures measured the
thermal emission and scattered light from the telescope mirror cover,
and we moved the telescope to different positions because the signal
instabilities seemed to be a function of telescope position, as
described in Sec.~3.2.

During our WASP-33 oservations, we also noticed relatively prominent
intensity oscillations of the star in $K_s$-band.  Hence, we also
observed 651 consecutives exposures of WASP-33 on the night of UT 14
October, when no eclipse (or transit) of the planet occurred.  This
time series was observed using J-band, and helps to establish the
degree to which the amplitude of stellar oscillations in intensity may
depend on wavelength.

\subsection{Warm Spitzer Observations}

Warm Spitzer observed one eclipse of WASP-33 at 3.6\,$\mu$m, for 9.6
hours on 26 March 2011, as well as observations of equal duration
spanning one eclipse at 4.5\,$\mu$m on 30 March 2011.  The
observations were made under the Cycle-6 Target of Opportunity program
(J. Harrington, P.I.).  Both observations used subarray mode to
collect 1264 data cubes, each data cube comprising 64 exposures of a
32$\times$32-pixel section of the detector.  The exposure time was 0.4
seconds per exposure, and the observations were not interrupted by any
re-pointing.

\section{Photometry}

\subsection{Ground-Based Photometry}

In both $K_s$ and J-band, we subtract a dark frame from each image,
and divide by the sky-flat for that wavelength.  Each sky flat is
normalized to have an average value near unity; this facilitates
conversion of the observed images from data numbers to electrons using
the known gain of the detector electronics (4.9 e/DN).  In the limit
of large defocus, the stars are essentially images of the telescope's
primary mirror; they are not well-approximated using Gaussians or
other functions commonly used for centroiding.  Therefore we determine
the center position of each star by exploiting the sharp edges that
are characteristic of the pupil images.  Summing the image of each
star in the X-coordinate produces intensity as a function of Y,
$I(Y)$.  The sharp edges of the pupil image will produce peaks in the
derivative $\partial{I}/\partial{Y}$, one peak at each edge of the
image. We find those peaks in the spatial derivative, and adopt an
average of the Y-positions of derivative peaks on each side of the
$I(Y)$ profile as being the center of the star in $Y$, and vice-versa
for $X$.  With the center of each star determined for each exposure,
we calculate the flux in a circular aperture of a given radius
centered on that star.

In addition to WASP-33, we measure 4- to 6 comparison stars on each
image.  The identification of the comparison stars are given in
Table~1, with their J and K magnitudes.  The subset of comparison
stars that were actually used varied from night to night, due to using
the J-band wavelength on 14 October, and K-band sky transparency and
thermal background that varied between 10 and 16 October. The subset
of comparison stars used for each night are noted in the caption of
Table~1.

For both the target and comparison stars, we determined the value of
the sky background by constructing a histogram of pixel values in a
$100 \times 100$-pixel box surrounding each star.  The sky background
dominates the peak of those histograms, and we fit a Gaussian to each
histogram, and thereby determine the sky background value for each
star as the centroid of each Gaussian.  Multiplying the background
value per-pixel for each star, times the area of the aperture
containing that star, yields the background contribution for that
star.  Subtracting the background from each aperture measurement
yields the stellar intensities of WASP-33 and the comparison stars for
that image.

We repeat our photometry by varying the radius of the synthetic
aperture from 18 to 30 pixels, in 1-pixel increments.  The 13
different aperture radii produce 13 different realizations of
photometry for each night.  Our rationale for varying the aperture
radius is to optimize the signal-to-noise and the match between the
target star and the comparison stars.  Although our images are
strongly defocused, there is still scattered light beyond the edges of
the defocused pupil image for each star. Variable seeing and errors in
aperture centering cause the total flux intercepted by a chosen
aperture to vary.  Measurements with larger apertures are less
sensitive to these variations, but include more background noise.  The
optimum aperture radius is approximately 20 pixels, but we determine
it separately for each night, as described in Sec.~4.1.

\subsection{Instrumental Instabilities}

As noted in the Introduction, we experienced instabilities in the
photometric signals.  These instabilities were discovered using
quick-look evaluation of the data during each night of observations.
The nature of the instabilities is that sharp increases or
decreases in stellar intensities occur typically 4 to 6 times each
night. These changes in intensity are large compared to our
photometric precision: sudden changes as large as $4\%$ were seen.
Stars close together in angular extent experienced similar, but not
identical effects.  For our observations, the comparison stars were
typically several hundred pixels distant from WASP-33, so the
instabilities were not precisely common-mode.  The sky background also exhibited
this instability, but to a much smaller relative degree - barely detectable.

During our observing run, we noticed that the signal instabilities
never occurred for stars at negative declination, and they tended not
to occur at large hour angles.  We concluded that these puzzling
instabilities were triggered by motion of the telescope, and that
motivated our diagnostic observations described in Sec.~2.  The root
cause of the signal instabilities is now known: following our
observing run, Dick Joyce of the Kitt Peak scientific staff
disassembled FLAMINGOS, and discovered that a 4-40 screw (probably
from the filter box) had come loose and fallen onto the first camera
lens, producing field-dependent vignetting as it shifted position.  We
were initially tempted to discard all of these `screwy' data, but the
excellent photometric quality of the nights (especially on 10 October)
motivated us to develop a methodology to remove the effect of the
rolling screw.

Fortunately, the nature of this effect - a sudden shift in position of
the screw followed by periods of stability - makes it amenable to
robust correction using only the data themselves.  We used our test
observations with the dome closed to validate our correction
methodology.  We produce aperture photometry from these frames using
apertures of 20-pixel radius, and no background subtraction (the
background {\it is} the signal for the test observations).  We center
the apertures at the same locations as WASP-33 and the comparison
stars, and extract time series signals.  Figure~1 (top panel) shows
the time series at the WASP-33 position.  The large and sudden changes
in signal level are obvious, and they correspond to times of telescope
motion.  The telescope motion for these test observations was done in
a series of 0.5-hour increments, versus continuous tracking for the
actual stellar observations.  Nevertheless, the results are very
similar, and we conclude that we have successfully reproduced the
signal instabilities.

We correct these instabilities by operating on the time derivative of
the signal. We calculate the distribution of the ensemble of the
numerical time derivative values for each time series.  This
distribution is primarily Gaussian, with extreme outliers that
correspond to the times of signal instability.  We fit to the Gaussian
core of the distributions, to determine the unbiased standard
deviation of the signal derivative at the position of each star.
We adopt a threshold value (typically $3\sigma$) and we correct the
outlying points in each series of derivative values, where the
absolute value of the derivative exceeds this threshold.  This
threshold value (in $\sigma$) is an adjustable parameter in our
correction procedure.  After identifying the derivative points beyond
the threshold, one option is to set these outlying derivative points
to zero.  However, in practice we replace derivative outliers with a
5-minute smoothed version of the derivative.  We choose the 5-minute
smoothing time because it represents the typical time for significant
changes in stellar intensity, e.g., as caused by the telluric
atmosphere.  After correcting the outlying points in the derivatives,
we integrate the corrected derivatives to yield corrected time series.

Figure~1 (upper panel) shows the corrected time series for our dome
test observations at the WASP-33 position. The derivative of the time
series is shown in the middle panel, with the rejected outliers
marked. The lower panel of Figure~1 shows the result of dividing the
corrected time series at the WASP-33 position, by the total of the
corrected series at the comparison star positions.  In the absence of
our correction procedure this quotient would show considerable noise
at the several-percent level.  The correction procedure successfully
produces a quotient whose short-term variations are of order 200
parts-per-million (ppm), with long-term variations approaching 700
ppm.  We do not regard the long-term variations as meaningful to our
correction procedure, because the background in the telescope dome is
not guaranteed to be stable at this level.  We conclude that our
procedure to correct for the signal instabilites is effective, and we
apply it to our stellar photometry.

\subsection{Warm Spitzer Photometry}

Photometry of Warm Spitzer data is now a familiar exercise
\citep{hebrard, beerer, deming11a, demory11, desert11, todorov}, and
we used well-tested procedures applied to the BCD files produced by
version S18.18.0 of the Spitzer pipeline.  We performed aperture
photometry on each frame, in each data cube, for both eclipses.  We
found the centroid of the stellar image by fitting a 2-D Gaussian, and
computed the flux in a circular aperture centered on the star, for
aperture radii from 2.0 to 5.0 pixels in 0.5-pixel increments.  We
accounted for the background contribution in each frame; the per-pixel
background was measured as the centroid of a Gaussian fit to a
histogram of pixel values for that frame.  Multiplying the background
per-pixel times the area of the aperture yielded the background
contribution that was subtracted from the flux in the aperture.
Aperture radii near 3.0-pixels produced the highest signal-to-noise
photometry, as judged by the point-to-point scatter. Accordingly, we
adopted a 3.0-pixel radius for our photometry at both 3.6- and
4.5\,$\mu$m.

\section{Analysis of the Photometry}

\subsection{Ground-based Photometric Analysis}

\subsubsection{Optimization of Photometric Parameters}

As described in Sec.~3.1, we generated 13 versions of our ground-based
photometry using a range of aperture radii, and we also have a free
parameter (the threshold) to correct the photometry for instrumental
instabilities.  We determine which combination of aperture radius and
correction threshold produces the most robust results.  We increment
the aperture radius in 1-pixel steps, and the correction threshold in
steps of $0.1\sigma$.  For each combination of aperture radius and
correction threshold, we calculate the linear Pearson correlation
coefficients between the photometric time series for WASP-33, and the
corresponding photometry for each comparison star.  In the limit of
very high correction threshold (i.e., no instability correction),
there is very little correlation between the comparison stars and
WASP-33.  That is consistent with the behavior of the instrumental
instability, as evaluated in our closed-dome test observations,
wherein we found that different portions of the detector exhibit
different instabilities.  Similarly, we expect photometric aperture
radii that are too small or too large would degrade the correlation
between WASP-33 and the comparison stars.

In order to select the best combination of aperture radius and
correction threshold, we average the correlation coefficients that we
compute for the comparison stars versus WASP-33.  For our J-band
observations on 14 October, the highest average correlation
coefficient ($0.85$) is achieved when the combination of (radius,
threshold) is (22 pixels, $3.1\sigma$).  For $K_s$-band observations
on 10 and 16 October, the optimum combinations are (22, $2.8\sigma$)
and (20, $3.1\sigma$), producing average correlation coefficients of
$0.88$ in both cases.  These results are quite reasonable, because
aperture radii near 20-pixels are modestly greater than the radius to
the half-intensity point of the defocused image (about 15 pixels), and
20-pixels is what our intuition told us to choose for quick-look
analyses at the telescope.  Similarly, threshold values near $3\sigma$
are reasonable because they allow sudden spikes in the signal
derivatives to be identified without perturbing normal fluctuations
due to photometric noise.  We conclude that the resultant time series
photometry is the best that can be achieved from these data.  As a
by-product of this process, we evaluate the point-to-point
fluctuations in the WASP-33 photometry, from the fit to the Gaussian
distribution of signal derivative values.  These fluctuations are
sub-milli-magnitude for all three nights; we find J-band noise of 642
ppm on 14 October, and $K_s$-band noise of 725 and 905 ppm on 10 and
16 October, respectively.  The higher noise on 16 October is produced
by a higher thermal background, due to warmer weather and slightly
degraded (but still excellent) atmospheric transparency over Kitt
Peak.

\subsubsection{Normalization using Comparison Stars}

Following selection of the best photometric aperture radius, and
threshold for instrumental correction, we further correct the WASP-33
time series using the comparison stars.  Normally this would be
accomplished by dividing WASP-33 by the sum of the comparison stars.
However, we find improved results using a slightly different
procedure: instead of a straight sum of the comparison stars, we use
a weighted sum. We denote the intensity of WASP-33 at time index $i$
as $W_i$, and we write:

\begin{equation}
W_i = \sum_{j=1}^{N} \alpha_j c_{ij},
\end{equation}

where $c_{ij}$ is the intensity of the $j$-th comparison star at time
index $i$, and the $\alpha_j$ coefficients - one for each of the $N$
comparison stars - are determined by linear regression (matrix
inversion).  The linear regression seeks to produce the best overall
match between the weighted sum of the comparison stars and WASP-33,
i.e., to make Eq. (1) be exact.  Because of noise and intrinsic
variations in WASP-33, Eq.(1) can never be solved exactly, only for
the set of $\alpha_j$ that produces the best approximation.  Having
solved for those $\alpha_j$, we divide the $W_i$ by the right hand
side of (1).  This division by the weighted sum of the comparison
stars removes instrumental and telluric effects, but leaves noise and
the intrinsic variations of WASP-33 itself.   

Eq.(1) is a generalization of the usual methodology of ratioing the
target star to the sum of the comparison stars.  Using an
equally-weighted sum of the comparison stars can be an imperfect
divisor, for several reasons.  First, the comparison stars usually
differ in spectral type from the target star, and the integral of
their flux over our broad filter produces a slightly different
effective wavelength for each star.  The extinction in the stellar
signals caused by the terrestrial atmosphere can vary strongly with
wavelength, so different effective wavelengths can degrade the
correlations between WASP-33 and the comparison stars.  Second, the
comparison stars lie at quite different locations on the detector, and
higher order effects can be different at these different locations.
For example, our data exhibit a slow drift in the position of all
stars over the course of a night (about one arc-sec total), caused by
differential refraction between the optical guider and the infrared
wavelength used for imaging.  That small drift, combined with small
inaccuracies in the flat-field calibration, can produce different slow
variations for each comparison star.  Also, slight changes in the
degree of defocus can result from inaccurate compensation of thermal
effects and mechanical flexure (Sec.~2.1), and can produce subtle
changes in the defocused images - hence in the aperture photometry -
as a function of field position.  For these reasons, we believe it is
good practice to optimally-weight the comparison stars using linear
regression.  

We compared our optimally-weighted photometry to photometry that used
an unweighted sum of the comparison stars.  The point-to-point scatter in
the two photometric data sets was about the same, but the baseline was
much flatter using the optimal technique.  Specifically, we found that
the unweighted sum produced overall slopes of 0.6\% and 0.8\% for the
nights of 16 and 10 October, respectively.  Using the optimal
technique reduced these baseline slopes to 0.03\% on both nights.

The linear regression used to solve Eq. (1) can potentially affect the
measured eclipse, because the regression attempts to remove all
fluctuations in the target star, and the eclipse is a fluctuation.  In
the hypothetical case where one of the comparison stars happened to exhibit a
noise-like decrease at the expected time of the WASP-33 eclipse, the
regression would overweight that star and the result would be to weaken
or remove the eclipse.  We avoid that possibility by solving for the
$\alpha_{j}$ using only out-of-eclipse portion of the data (adopting the
central phase as 0.5, and setting the duration of eclipse to equal the
duration of the transit.) 

Figure~2 shows the photometry for WASP-33 in the $K_s$ band on both 10
and 16 October, and the weighted sum of the comparison stars (right
hand side of Eq. 1) is overplotted as a blue line.  The overall rise
and fall of these signals is due to the airmass dependence of telluric
absorption, and shorter term fluctuations due to telluric effects are
also visible.  Note also that the weighted sum of the comparison stars
shows more short-term noise than does WASP-33b, because most of the
comparison stars are 1- to 2-magnitudes fainter than WASP-33 (see
Table~1). For that reason, we smooth the comparison star sum over 10
points (about 7 minutes in time) before using it to correct WASP-33.
That degree of smoothing decreases the point-to-point noise in the
ratio, while still following most telluric fluctuations.

\subsection{Warm Spitzer Photometric Analysis}

A dominant effect in Spitzer photometry at both 3.6- and 4.5\,$\mu$m
is the presence of intra-pixel sensitivity variations. The photometric
intensity of a star will depend on its position on the detector, and
therefore will vary with time because of a pointing oscillation in the
telescope (\citealp{carey}; see \citealp{todorov} for a recent
example).  The Spitzer project recently implemented software updates
that decrease the amplitude of the pointing oscillation, and also
decrease its period from 1-hour to about 40
minutes\footnote{http://ssc.spitzer.caltech.edu/warmmission/news/21oct210memo.pdf}.
This reduces the impact of the intra-pixel sensitivity variations on
the photometry.  Indeed, our observations of WASP-33 show the
intra-pixel effect to a much less degree than many previous
observations.

Many previous investigations have established that the intrapixel
effect is more dependent on the Y-coordinate than on the X-coordinate,
and it is stronger at 3.6- than at 4.5\,$\mu$m \citep{knutson09,
beerer, deming11a, todorov}.  We see the intrapixel effect in our
3.6\,$\mu$m photometry, and we corrected it using a quadratic fit to
the photometry as a function of the Y-coordinate of the image, with a
linear term as a function of the X-coordinate (the variation with X is
weaker than for Y).  However, we cannot detect any intra-pixel
sensitivity variations in our 4.5\,$\mu$m photometry.  Plots of the
measured intensity of the star versus both the X- and Y-coordinate of
the image centroid are essentially scatter plots (not illustrated),
with no signficant trends.  Pearson correlation coefficients of
intensity versus X and Y-coordinate have values near zero.  We also
calculated the Pearson correlation coefficients for temporal subsets
of the 4.5\,$\mu$m data, chosen using a moving boxcar window of
various widths. We can find no significant correlations between our
4.5\,$\mu$m photometry and spatial coordinates, during any time
period. As an additional check, we repeated our photometry using an
alternative method to determine the position of the star on the
detector (center-of-light). This method also failed to reveal
correlation between position and intensity.  We therefore use our
4.5\,$\mu$m photometry for eclipse analysis without applying any
spatial decorrelation.  Note that we do see very small variations in
the photometry at the 40-minute period corresponding to the telescope
oscillation (see Sec.~4.4.2). We cannot rule out the possibility that
the spatial-intensity correlation is being obscured by the stellar
intensity oscillations.

In the case of the 3.6\,$\mu$m photometry, we also used the spatial
decorrelation method described by \citet{ballard}.  That weighting
function method commonly uses a time threshold to zero-weight points
that lie close in time to any given point.  When the only intrinsic
temporal variation is an eclipse or transit, the time threshold is
straight forward to implement.  However, when continuous stellar
oscillations are part of the desired signal, the weighting function
time threshold can be problematic. We applied a weighting function to
the 3.6\,$\mu$m photometry, using a time threshold of zero, i.e. not
excluding any points in the calculated weights.  Fitting to this
alternative version of the 3.6\,$\mu$m photometry produced results
that differed insignificantly from our quoted results ($0.34\sigma$,
and $0.03\sigma$ differences in central phase and eclipse depth,
respectively).  Our quoted results (see Sec.~4.4.2) are based on the
polynominal decorrelation described above, because we believe that
method is better suited to the nature of these data.

After decorrelating (or not) the intra-pixel effect, we omit some of
the initial data for both Spitzer wavelengths.  The 3.6\,$\mu$m data
exhibit an initial transient drift in the image position, amounting to
about 0.17-pixels, about 4 times as large as the peak-to-peak
variation caused by the 40-min telescope oscillation. This positional
instability is accompanied by a similar large transient effect in the
photometry.  We are familiar with the nature of these large transient
effects based on seeing them in other Spitzer data.  The relation
between intensity and X,Y position is not the same during the
transient as during the stable portion of the time series, because the
image traces a different region of the pixel.  Therefore, we omit the
first 83 minutes of the 3.6\,$\mu$m data - this being the time for the
position of the image to stabilize.  We see no obvious transient
effects in the 4.5\,$\mu$m image position, but the first 27 minutes of
the photometry are anomalously low.  As a precaution, we omit those
initial data from our 4.5\,$\mu$m analysis.

\subsection{Observed Stellar Oscillations}

Figures 3, 4, and 5 show the ground-based WASP-33 photometry after all
corrections.  WASP-33 is known to be an oscillating $\delta$-Scuti
star \citep{collier-cameron, herrero}.  \citet{herrero} find a
dominant oscillation period of 68 minutes, but oscillating stars can
exhibit different oscillation modes that rise and fall in
amplitude. We have therefore decomposed the photometry on each night
using wavelets \citep{torrence} to account for possible intra-night
changes in the oscillatory spectrum.  Because the eclipse can
significantly affect the wavelet analysis, we remove the eclipse prior
to computing the wavelet power spectra. For the removal, we adopt the
best-fit eclipse parameters as derived in Sec.~4.4.  (This is not a
circular procedure because the fits for the eclipse parameters
converge relatively independently of the starting estimates for the
oscillation periods.)

Our results show the stellar oscillations very prominently, but
(interestingly) to varying degrees.  The J-band results are shown in
Figure~3; the bottom panel shows the corrected photometry and the
overlaid blue line shows part of the wavelet decomposition - just the
lowest order components to guide the eye to the general trend.  (Note
that we do not use the blue lines on Figures 3 - 5 in our analysis,
but we do use the results of the wavelet power spectra, shown in the upper panels.)

The power spectrum in Figure~3 peaks at 146 minutes (frequency 9.9
cy/d), close to twice the period of the 68-minute oscillation (21.2
cy/d) seen by \cite{herrero}.  No mention of this lower frequency mode
was reported by \citet{herrero}, but $\delta$-Scuti stars are known to
exhibit multiple periods of oscillation \citep{balona}, wherein the
non-radial modes cluster around different radial modes (e.g.,
$n=1,2,3...$, \citealp{breger09}). Hence, detection of a different
mode frequency than \citet{herrero} is not surprising.  Figure~4 shows
the results in similar format for the $K_s$-band observations on
10~October.  In this case a very prominent oscillation is seen with a
power spectral peak at 71 minutes (20.3 cy/d), in good agreement with
essentially the same oscillation period (68.56 minutes) that was
prominent in the results of \citet{herrero}. Note also that it can be
clearly discerned in the photometry even prior to the correction using
the comparison stars - see Figure 2. The amplitude of this oscillation
- about $0.2\%$ in flux, is about twice what was observed by
\citet{herrero} in the optical, and we discuss possible reasons for
this difference in Sec.~5.

Figure~5 shows the analogous result for the $K_s$ band observed on
16~October.  In this case we see two peaks in the power spectrum at 54
and 126 minutes, that are probably due to the stellar oscillations as
seen on 10 and 14~October.  For each night, and also for the Spitzer
data, we estimate the total stellar oscillation amplitude from the
power present in the oscillatory component of the MCMC eclipse fit
(see Sec.~4.4), and we tabulate those amplitudes in Table~2.  All of
these amplitudes are in reasonable agreement with the 0.1\% oscillation seen
in the optical by \citet{herrero}, except in $K_s$ band where the
amplitude is about twice as great, consistently on both 10 and 16
October.  

Variability in stellar oscillation amplitudes is sometimes observed in
$\delta$-Scuti stars (e.g., \citealp{breger06}), but is not the most
likely explanation for our results, because of our J-band data.
Although J-band shows a prominent oscillation near phase 0.82 on
Figure~3, quantitative calculation of the average amplitude over that
entire night gives 0.16\% (Table~2), only marginally higher than the
optical value.  A close-to-normal oscillation amplitude in J-band,
between the two nights of higher amplitude in $K_s$ band, seems too
much of a coincidence to attribute to variability.  While we cannot
strictly reject mode variability, we hypothesize that oscillation
amplitude is associated with wavelength. We discuss possible wavelength-related
explanations in Sec.~5.2.

\subsection{Eclipse of WASP-33b}

\subsubsection{$K_s$-band Eclipse}

The eclipse of the planet is visible near phase 0.5 in the lower
panels of Figures~4 and~5.  No eclipse is seen in Figure~3, as
expected for that range of orbital phase.  The stellar oscillations
are a very significant source of confusion as regards measuring the
properties of the eclipse.  That confusion is mitigated by
sufficiently long duration of our observed sequences (7.3 hours on 10
October amd 5.4 hours on 16 October) compared to the periods of
oscillation, and by the fact that the stellar oscillations will not be
in phase on the two nights, whereas the eclipse should repeat at the
same orbital phase.  In order to exploit the latter advantage, we
combine the data for 10 and 16~October into one binned data set, using
a bin width of 0.001 in phase, and we fit models to these binned data.
Figure~6 shows the combined and binned data, with best-fit eclipse
plus oscillation curves overplotted.

To obtain the best-fit eclipse depth, we fit a combination of eclipse
curve plus two sinsusoidal oscillations, using a Markov Chain Monte
Carlo (MCMC) method, with Gibbs sampling \citep{ford}. We fit to the combined
and binned data of Figure~6, in order to gain the advantage of
canceling the stellar oscillation as much as possible. We generate
four chains of $10^6$ steps each, and we discard the first $2 \times
10^5$ steps in each chain.  We initialize the MCMC fit using sinusoids
with periods of 68 and 146 minutes, guided by our wavelet power spectra
in Figures~3--5.  The MCMC fit is insensitive to the exact choice of
initital periods.  As long as the initial values contain one period in
the $\sim 70$-minute range, and one period in the $\sim 140$-minute
range, the Markov chains rapidly converge to the best-fit values for
that particular eclipse (this is also true for the Spitzer eclipses,
Sec. 4.4.2).

In order to derive accurate MCMC posterior distributions, it is
essential to re-scale the error bars to yield a best reduced $\chi^2$
near unity.  We find that errors about four times larger than the
photon noise (for the binned data) are necesary to obtain a best-fit
reduced $\chi^2$ of unity.  Error rescaling factors almost as large as 3 have
been required in other investigations, even for very precise
spectrophotometry \citep{bean}.  An even larger error re-scaling in
our case is probably a consequence of the complexity of the stellar
oscillation spectrum, that is only approximated using two oscillation
periods.  In other words, small amplitude residual oscillations may
masquerade as noise.

We experimented with adding additional oscillation modes to the fit,
and indeed this does reduce the necessity for error rescaling.
However, we can only justify initializing the fit using the two
periods that we can objectively identify in our wavelet power spectra.
Other modes, if present, cannot be resolved as clear oscillations in
our ground-based data, instead they appear only as extra noise.  We
also verified that adding a third oscillation mode to our MCMC fit
does not significantly alter the best-fit eclipse depth.

We vary nine parameters in the MCMC chains: an amplitude, period, and
phase for two independent sinusoids, an eclipse depth and central
phase, and an additive constant.  We generate the eclipse curve using
a new version of the \cite{mandel} methodology (see
\citealp{deming11b}). We adopt system parameters, that determine the
shape and duration of the eclipse, from \cite{collier-cameron}, except
for the orbital period where we use the value from \citet{smith}.
Observations of recent transits in the Czech Exoplanet Transit
Database\footnote{http://var2.astro.cz/ETD/} show that the original
period from \citet{collier-cameron} is too short; the slightly longer
period derived by \citet{smith} is more consistent with recent transit
times in the Czech database.

The top panel of Figure~6 overplots the best-fit curve, including both
the eclipse and oscillatory component.  The lower panel removes the
oscillatory component from the data, and shows the comparison with the
eclipse curve alone.  Our best-fit eclipse has a depth of
$0.0027\pm0.0002$, with central phase of $0.4995\pm0.0010$.  Those
errors are implied by our MCMC posterior distributions.  They include
possible degeneracies between fitted parameters, but there can be
external sources of error than are not represented in the MCMC chains.
The most obvious source of such errors is the presence of unmodeled
stellar oscillation structure, as noted above.  The most realistic
check on the errors is to fit each of the two independent nights
separately, and compare the independent results.  Those fits yield
eclipse depths of $0.0024\pm0.0002$ and $0.0033\pm0.0004$ for 10 and
16~October, respectively.  The best-fit central phases for the
separate nights are $0.502\pm0.001$ and $0.495\pm0.001$, so the
difference between two independent nights exceeds the formal errors,
especially in the case of central phase.  We adopt the best-fit values
determined for the combined and binned data, but we assign the errors
based on the differences in the two independent nights.  The
difference between two independent values drawn from a normal error
distribution is, on average, twice the error associated with the
average of those two values.  So we estimate the errors appropriate to
our average $K_s$-band eclipse depth and central phase to be half the
difference between the best-fit values for the individual nights.  Our
best-fit eclipse depth and central phase is given in Table~3, together
with the Spitzer eclipse results.

\subsubsection{The Eclipse in Warm Spitzer Bands}

At both Spitzer wavelengths, we solve for the eclipse depth and
central phase also using an MCMC method, with Gibbs sampling.  We
allow for an exponential baseline ramp \citep{harrington}, because we
find that a linear ramp is inadequate.  Indeed, detailed analysis of
very high S/N Spitzer 8\,$\mu$m data \citep{agol} indicates that even
a second exponential term can be warranted.  However, we use a single
exponential for two reasons.  First, the single exponential ramp alone
is more complex than is normally required for these specific Spitzer
bands \citep{knutson09, todorov} - exponential ramps are normally only
required at the longest wavelength Spitzer bands.  Second, our data do
not attain sufficient S/N to justify the inclusion of a second
exponential term.

As in the case of our ground-based data, we calculate wavelelet power
spectra at both Spitzer bands, and these are shown as the upper panels
of Figures~7 and~8.  Like the ground-based data (perhaps a
coincidence), we see two significant oscillatory spectral peaks after
removing the best-fit eclipse.  Figure~7 (3.6\,$\mu$m) shows one of
these peaks at 68 minutes. However, Figure~8 (4.5\,$\mu$m) shows two
peaks, 39 and 68 minutes.  The former is likely to be associated with
the Spitzer telescope pointing (Sec.~4.2), while the latter is clearly
due to the stellar oscillation.

Our first MCMC fits at both Spitzer wavelengths vary 12 parameters:
two sinusoidal terms (each having amplitude, period and phase), an
eclipse of variable depth and central phase (i.e., two parameters),
the exponential ramp (three parameters), and a zero-point constant in
intensity.  We discard the first 20\% of each $10^6$-step Markov chain
and locate the best fit values from the minimum in $\chi^2$ (we keep
track of $\chi^2$ during the evolution of the chain).  Subtracting the
best fit from the data, we find that the residuals have a
quasi-sinusoidal shape. These residuals represent the unmodeled
portion of the data.  They can be due to red noise in the data created
by uncorrected instrumental effects, or by imperfect modeling of the
stellar oscillations.  Unlike the case of our ground-based data, the
Spitzer data have sufficient stability and S/N to define these small
amplitude imperfections in the fit. Both the oscillating intensity of
the star, and (to a lesser extent) the subtlety of red noise caused by
Spitzer's improved performance, require some new methodology in their
analysis.  We now introduce that new methodology.

Following the inital fit using a single MCMC chain of $10^6$ steps, we re-fit
by including an additional term to explicitly account for the
imperfections in the initial fit. We calculate residuals for the first
MCMC fit by subtracting that best fit from the data.  We then
approximate those residuals using a wavelet decomposition, using $N$
coefficients for Morelet wavelets, and we vary $N$ in our subsequent
analysis.  For each choice of $N$, we multiply the wavelet
decomposition of the residuals by an adjustable amplitude, and include
that amplitude as a fit parameter in subsequent MCMC chains.  In the
limit where $N$ equals the number of data points, this procedure would
be trivial, because the wavelet decomposition would reproduce the
residuals exactly, and including those residuals in the fit would
simply be a fudge, wherein we arbitrarily subtract that part of the
data not accounted for by the model.  In the real analysis, this
procedure is not trivial.  With not-too-large $N$, the wavelet
decomposition approximates only the major features of the residuals,
not the point-to-point noise.  Because the MCMC chain varies the
amplitude of those major features, correlating the chained values of
eclipse depth and phase with the amplitude of the unmodeled features
indicates the degree to which the best-fit eclipse depth and phase
depend on unmodeled aspects of the data.  We determine the optimal $N$
by examining the noise in the final fits, requiring that it be close
to, but not less than the photon noise, and have no detectable red
noise component.  This dictates $N=9$ for both our 3.6\ and
4.5\,$\mu$m data. Our final values are based on 4 independent MCMC
chains at $N=9$, each having $10^6$ steps and ignoring the first 20\%
of each chain.

Using this procedure, we find that the Spitzer eclipse depths and
central phases are reasonably robust in the sense that they do not
vary greatly with $N$. This is especially true for the central phase
at 3.6\,$\mu$m, and the eclipse depth at 4.5\,$\mu$m.  In those cases,
the variation in eclipse depth and central phase, as we vary $N$, are
consistent with the errors implied by the MCMC posterior
distributions.  However, for the converse set of parameters - the
3.6\,$\mu$m eclipse depth and 4.5\,$\mu$m central phase - the
agreement is not as good. In those cases, the differences in best-fit
values as we vary $N$ are larger than the errors from the MCMC
posterior distributions, and we adopt the larger errors consistent with the
fluctuations in the best-fit values as $N$ varies.

The derived best-fit eclipse depths, central phases, and errors are
included in Table~3.  We also extract the amplitude of the stellar
oscillation in the Spitzer bands (see Table~2), based on the total
amplitude of the oscillatory portion of the MCMC fits, but with the
40-minute portion subtracted because we attribute that portion to the
Spitzer observatory.

\subsection{Models for the Atmosphere of WASP-33b}

We interpret our observed planet-star flux contrast of WASP-33b using
plane-parallel models of the dayside atmosphere of the planet.  We use
the atmospheric modeling and retrieval methodology of \citet{madhu09,
madhu10}. The model computes line-by-line radiative transfer for a
plane-parallel atmosphere with the assumptions of hydrostatic
equilibrium and global energy balance. The composition and
pressure-temperature ($P$-$T$) profile of the atmosphere are free
parameters in the model. Since there are as yet insufficient data to
constrain abundances in WASP-33b, we adopt solar composition for all
elements except carbon, which we allow to increase in abundance,
following \citet{madhu11b}.  We explore a variety of temperature
profiles, especially temperature profiles that are consistent with
nearly complete absorption of stellar irradiance.  The models include
all the major opacity sources expected in hot hydrogen-dominated
atmospheres, namely H$_2$O, CO, CH$_4$, CO$_2$, TiO, and VO, and
collision-induced absorption (CIA) due to H$_2$-H$_2$. Our molecular
line lists are obtained from \citealp{freedman08},
\citealp{freedman09}, \citealp{rothman05}, \citealp{karkoschka10} and
\citealp{karkoschka11}. Our CIA opacities are obtained from
\citet{borysow97} and \citet{borysow02}. A Kurucz model \citep{kurucz}
is used for the stellar spectrum, and the stellar and planetary
parameters are adopted from \citet{collier-cameron}.

\section{Results and Discussion}

\subsection{Stellar Oscillations}

The first and most obvious result from our observations is the
existence and prominence of the stellar oscillations.  WASP-33 was
already known to exhibit oscillations, but the amplitude observed by
\citet{herrero} in the optical (Johnson R-band) was about 0.001.  Our
results (Figures 3-5, \& Figures 7-8) show oscillation amplitudes in
agreement with the optical, except for $K_s$ band where the ampitude
is about twice the optical value (2.15\,$\mu$m, Table~2), as noted in
Sec.~4.3.  Because the largest difference with the optical amplitude
is seen in our ground-based ($K_s$-band) data, we contemplated whether
the difference could be attributed to errors in our ground-based
result. An argument against that possibility is the prominence of the
stellar oscillation in the raw photometry (e.g., upper panel of Figure~2).
We therefore explore whether properties of the stellar atmosphere that
may be unique to $K_s$ band could cause the oscillations to have
greater amplitude at that wavelength.

\subsection{Stellar Atmospheric Effects}

We here consider the possibility that the larger $K_s$ band
oscillation amplitude as compared to the optical is due to the
different height of formation for continuum radiation in the stellar
atmosphere, in concert with height-dependent variations in the mode
amplitudes. Due to the increase in atomic hydrogen bound-free
continuous opacity in the infrared, our $K_s$ band observations of WASP-33 sample
a greater height in the stellar atmosphere than observed by
\citet{herrero}.

The upward propagation of a pressure-mode oscillation in a stellar
atmosphere can in principle cause the mode amplitude to increase.  As
a propagating mode encounters lower mass density, the wave velocity -
and hence the temperature perturbation in the compression - increase.
However, propagation is strongly affected by the stratification of the
stellar atmosphere \citep{marmolino}.  Frequencies less than the
acoustic cutoff frequency will not propagate, and their velocity
amplitude decreases with height. To the extent that the temperature
amplitude scales with velocity, it too will decrease with height for
non-propagating modes. The acoustic cutoff frequency is $c/2H$, where
$c$ is sound speed and $H$ is the pressure scale height.  We
calculated the acoustic cutoff frequency and other parameters for
WASP-33, using a Teff/log(g)/[M/H] = 7500/4.5/0.0 model atmosphere
from \citet{kurucz}.  We find that the acoustic cutoff corresponds to
an oscillatory period of about 1 minute.  The much longer period
oscillations we observe for WASP-33 are therefore not propagating, and
their amplitudes should decay with height.

The dominant opacity due to atomic hydrogen \citep{menzel} is higher
at $2.15\,\mu$m versus the optical by a factor of about 1.6,
translating to a height difference of about 60 km.  That is not
sufficient to account for significant changes in the mode properties,
even for propagating modes and, as noted above, these modes do not
propagate.  Moreover, if height dependence of the stellar continuous
spectrum were significant to our observed amplitudes, then we would
expect even larger amplitudes in the Spitzer bands than at
2.15\,$\mu$m, which are not observed.

However, there is one unique feature of the $K_s$ band that may well
be responsible for a higher oscillation mode amplitude.  The
Brackett-$\gamma$ line at 2.165\,$\mu$m is centered in the $K_s$
bandpass.  The strong opacity in that line for an A5V star could have
a large potential effect on oscillation amplitudes.  The impact of
strong oscillations in the line, when diluted over the broad $K_s$
band, could potentially be calculated using techniques beyond the
scope of this paper (i.e., radiation hydrodynamics).  A more direct
method would be to obtain infared spectroscopy of the star and
directly measure the oscillations in the infrared hydrogen lines.

\subsection{Orbit of WASP-33b}

Our measured times of central eclipse can in principle determine
$e\,cos\,\omega$ for the planet's orbit (e.g., \citealp{knutson09}).
Considering the 25 seconds of light travel time across the planet's
orbit, we expect to find the eclipse at phase $0.50024$ if the orbit
is circular.  Weighting both our ground-based and Spitzer eclipses
(Table~2) by the inverse of their formal variances, we find an average
eclipse phase of $0.50044\pm0.00008$, totally dominated by the
3.6\,$\mu$m eclipse.  Thus we find that $e\,cos\,\omega$ - approximated
as $\pi/2$ times the phase offset from 0.5 - is $0.0003\pm0.00013$.
The central phase measurement for these eclipses is complicated by the
stellar oscillations, so more-than-usual caution is needed in the
interpretation of the measured central phase.  Moreover, our value for
$e\,cos\,\omega$ differs from zero by less than $3\sigma$, so our
results provide little evidence for a non-circular orbit.

\subsection{Atmosphere of WASP-33b}

Figure~9 shows our result for the eclipse of WASP-33b in comparison to
two models of its atmosphere, both of which agree with the available
measurements to date.  One of these models has a temperature inversion
with solar composition, and one has a non-inverted atmospheric
structure with a carbon-rich composition \citep{madhu11b}.  Their
temperature profiles are shown in the bottom panel of Figure~9, with
the approximate formation depths of the four bandpasses overplotted as
points. The $K_s$ bandpass is relatively devoid of strong molecular
absorption features, and probes the relatively deep planetary
atmosphere (pressure, $P$ $\sim$ 0.6 bars), relatively independent of
the composition of the atmosphere (asterisks on Figure~9). (The
3.6\,$\mu$m bandpass also peaks relatively deep in the atmosphere, but
has significant contribution from higher altitudes, having more
molecular absorption than does the $K_s$ band.)  The large eclipse
depth we observe in $K_s$ band (brightness temperature $\approx
3400$K) thus indicates a hot atmosphere at depth, and a high effective
temperature for the planet.  Both models illustrated on Figure~9 have
hot lower atmospheres (both above 2500 Kelvins) and both have
inefficient longitudinal energy re-distribution. These models are
consistent with the observed tendency for the most strongly irradiated
planets to exhibit the least longitudinal re-distribution of heat
\citep{cowan}. We find that these very hot models are necessary to
reproduce our results as well as the result of \citet{smith}, and we
conclude that WASP-33 strengthens the \citet{cowan} result.

The two models we show are representative of a larger set of solutions
that explain the data with and without thermal inversions. Given that
there are 10 model parameters \citep{madhu09, madhu10} and only four
data points, it is not possible to derive a unique model fit to the
data.  We ran large MCMC chains (of $\sim 10^6$ models) with and
without thermal inversions, and identified regions of composition
space in each case that are favored by the data \citep{madhu10}.

Both models on Figure~9 are unusual as compared for example to the
well-observed archetype HD\,189733b \citep{charb08}. One model on
Figure~9 adopts solar composition but with an inverted temperature
structure (temperature rising with height), while the other model has
temperature declining with height, but requires a carbon-rich
composition.  We integrate the fluxes of each planetary model, and the
Kurucz model stellar atmosphere, over the observational bandpasses,
and ratio those integrals.  These band-integrated points are shown as
squares on Figure~9.  The $\chi^2$ values for the models as compared
to all four observed points (our three measurements, plus
\citealp{smith}) are 7.9 for the inverted solar-composition model and
2.8 for the non-inverted carbon-rich model.  The $K_s$-band point at
2.15\,$\mu$m favors the non-inverted model. Although the difference is
not sufficiently significant to rule out the inverted model at this
time, additional eclipse observations in the $K_s$ band would be
helpful to rule out an inverted atmospheric structure.

In our population of models with thermal inversions, several models
with slightly different inverted temperature structure fit the data
almost equally well. However, none of them fit the K-band point to within the
1$\sigma$ errors while also fitting the remaining points. The
Figure~9 model is the best among this set of inverted models.

In our models without thermal inversions, the best-fit model requires
a carbon-rich composition (i.e., $C/O \geq 1$).  However, at the
2$\sigma$ level of significance per wavelength point, several solar
composition models (not illustrated) provide an acceptable fit to the
data.  So although the carbon-rich composition is favored, a solar
abundance composition cannot be absolutely ruled out.

Our results illustrate the limitations of eclipse photometry in broad
bands, especially for challenging cases like planets orbiting
oscillating stars.  Once we admit the possibility of non-solar
compositions (because we are largely ignorant of true exoplanetary
compositions), the range of models that can fit broad-band photometry
can be large, in this case extending to inverted and non-inverted
models with drastically different temperature structure. The
degeneracy is exacerbated by the relatively small range of atmospheric
pressures probed by the four bandpasses we analyze (points on the
bottom panel of Figure~9).  Fortunately, future observations can break
this degeneracy using HST/WFC3 spectroscopy near 1.4\,$\mu$m
\citep{berta12}.  The water band near 1.4\,$\mu$m is sufficiently
strong that eclipse observations with the Hubble WFC3 grism should be
feasible.  Although water absorption in the carbon-rich model is
supressed by $C/O > 1$, the water emission in the solar-composition
inverted model is predicted to be significant, readily detectable near
1.4\,$\mu$m (see Figure~9). Moreover, spectroscopic techniques can
potentially probe a larger depth range in exoplanetary atmospheres
than does photometry, because the cores of resolved spectral features
have strong opacities. Our results therefore illustrate the
complementary value of acquiring both broad-band and spectroscopic
observations of transiting exoplanets at secondary eclipse.

\section{Summary}

We have analyzed ground-based and Spitzer infrared observations of the
strongly irradiated exoplanet WASP-33b.  Our observations span the
time of secondary eclipse in $K_s$-band and the Warm Spitzer bands at
3.6- and 4.5\,$\mu$m.  We also observed the system for one
ground-based night out of eclipse in J-band (1.25\,$\mu$m).
Oscillations of the $\delta$-Scuti host star are prominent in our
data, with a semi-amplitude (0.1\%) about the same as in the optical.
One exception is the $K_s$-band, where we find an oscillation
amplitude about twice as large as seen in the optical.  Neither
temporal variability, nor the variation of stellar continuous opacity
with wavelength, are likely to account for our $K_s$-band result.  We
speculate that the greater amplitude in $K_s$-band may be related to the
presence of the Brackett-$\gamma$ line in the bandpass.

We measure two $K_s$ band eclipses, and we base our errors for the
eclipse depth and central phase on the difference between the two
independent measurements.  In the case of the Spitzer bands, we
measure one eclipse in each band.  Our Spitzer observations exhibit a
relatively low level of intra-pixel sensitivity variation as compared
to previous observations of other exoplanets.  However, all of our
observed eclipses are overlaid by the ubiquitous stellar oscillations.
We adopt an eclipse model comprised of the eclipse shape, plus two
sinusoids to account for the stellar oscillations.  We fit the model
to the photometry using an MCMC method.  The Spitzer photometry is of
sufficient quality that we are able to implement a wavelet-based
technique to define fluctuations in the data that are not accounted
for by our fitting procedure.  Including that structure as a parameter
in subsequent MCMC fits, we thereby explore how the eclipse depth and central
phase vary as a function of the amplitude of the unmodeled
structure. The derived eclipse depth and central phase of the Spitzer
eclipses are not strongly sensitive to unmodeled structure in the
light curves, but this procedure does allow a realistic evaluation of the errors.

Using the million-model approach pioneered by \citet{madhu09} and
\citet{madhu10}, we explore the range of atmospheric temperature
structures and compositions that are consistent with our eclipse
observations, plus the 0.9\,$\mu$m eclipse observed by \citet{smith}.
We find two possible atmospheric models.  One has an inverted
temperature structure, and solar composition, and the second has a
non-inverted temperature structure with a carbon-rich composition.  We
point out the value of infrared eclipse spectroscopy using moderate
resolving power to detect (for example) the water band near
1.4\,$\mu$m.  Spectroscopic detection of water emission or absorption
at eclipse would break the degeneracy between the two possible models.
Although the temperature structure of WASP-33b is currently ambiguous, both
models re-emit a large fraction of incident stellar irradiation from
their day sides, strengthening the hypothesis of \citet{cowan} that
the most strongly irradiated planets circulate energy to their night
sides with low efficiency.

\clearpage

%% the file, uncomment the \plotone, \plottwo, and
%% If you need a layout that cannot be achieved with \plotone or
%% \plottwo, you can invoke the graphicx package directly with the
%% \includegraphics command or use \plotfiddle. For more information,
%% please see the tutorial on "Using Electronic Art with AASTeX" in the
%% documentation section at the AASTeX Web site,
%% http://www.journals.uchicago.edu/AAS/AASTeX.
%%
%% The examples below also include sample markup for submission of
%% supplemental electronic materials. As always, be sure to check
%% the instructions to authors for the journal you are submitting to
%% for specific submissions guidelines as they vary from

%% This example uses \plotone to include an EPS file scaled to
%% 80% of its natural size with \epsscale. Its caption
%% has been written to indicate that additional figure parts will be
%% available in the electronic journal.

\renewcommand{\baselinestretch}{1.5}

\begin{figure}
\epsscale{.50}
\plotone{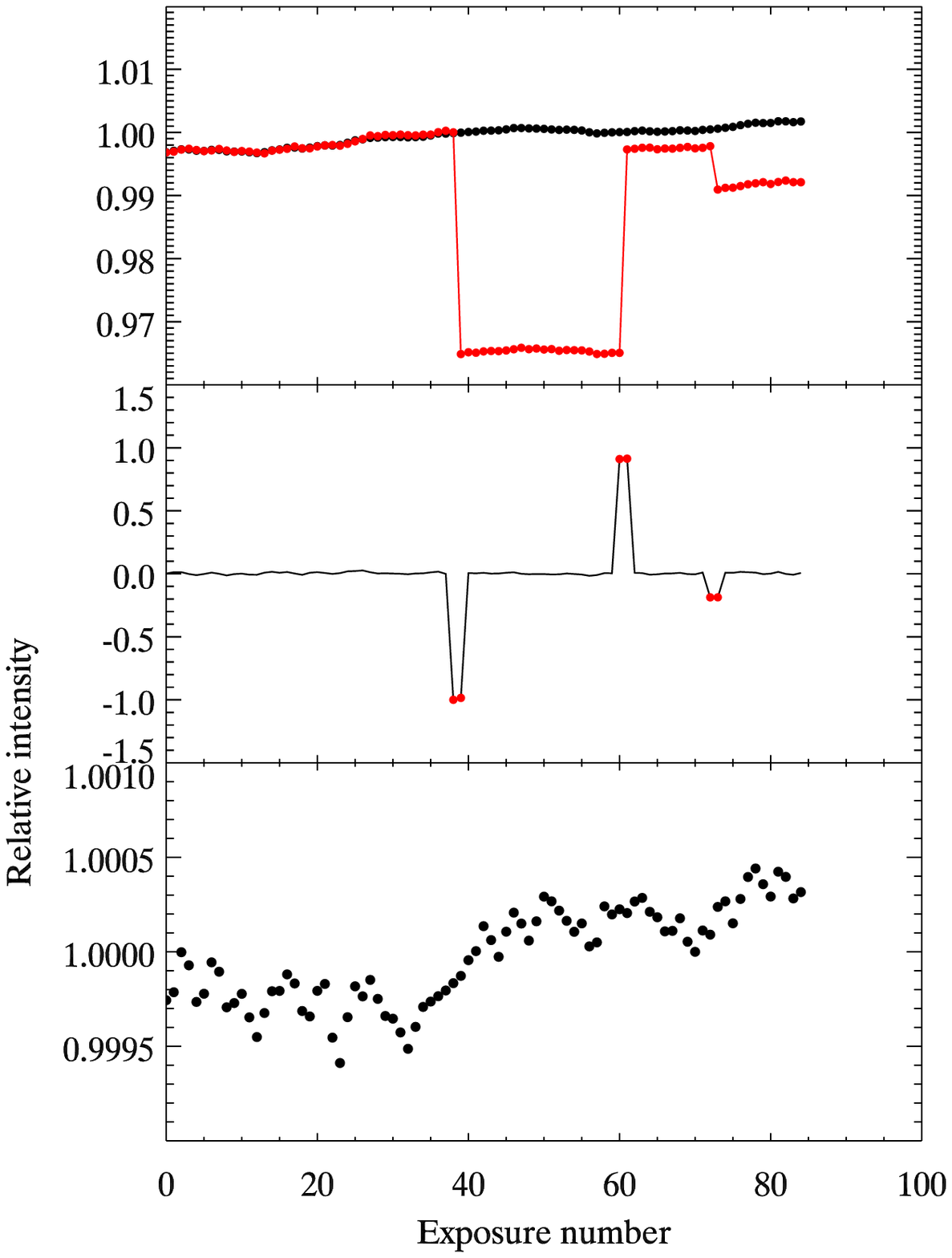}
\vspace{0.7in}
\caption{Results of our dome-test observations and procedure to
correct for the instrument-related signal instabilities.  {\it Top
panel:} original data (red points, connected with line) and corrected
data (black points) for our test observations at the WASP-33
position. {\it Middle panel:} Derivative of the original data from the
top panel, with the red points marking outliers in the derivative that are
corrected by our methodology.  {\it Bottom panel:} Ratio of the
corrected signal (black points) at the WASP-33 position to the sum
(not illustrated) of the corrected signals at the comparison star
positions.  Note the greatly expanded intensity scale on the lower
panel.  These test observations were made observing K-band background
with the telescope mirror and dome both closed.  The cadence of these
2-second exposures was about one per minute, with periodic telescope
motions in 0.5-hour increments of right ascension from -4 hours to +2
hours, at constant +32-degree declination.  The instabilities (e.g.,
at exposure 38) always correspond to times of telescope motion, but
not every telescope motion causes an instability.
\label{fig1}}
\end{figure}

\clearpage

\begin{figure}
\epsscale{.50}
\plotone{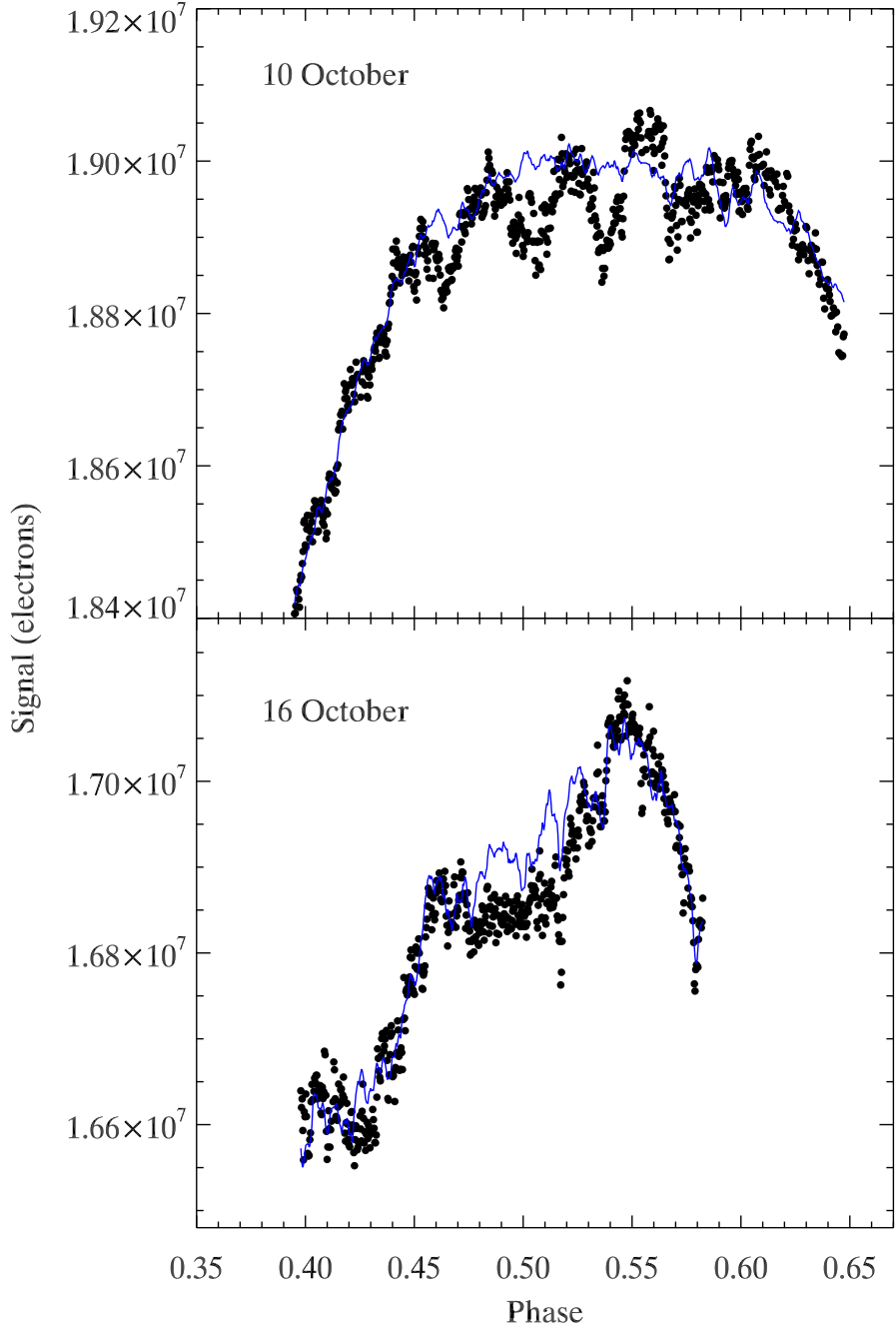}
\vspace{0.7in}
\caption{Photometry of WASP-33 (black points) in the $K_s$-band on 10
and 16 October, 2011, spanning the time of secondary eclipse. These
data have been corrected for instrumental instabilities, but not
normalized using the comparison stars. The blue line is the weighted sum
of the comparison stars, i.e., the right hand side of Eq. (1),
smoothed over 7 points (about 5 minutes of time).
\label{fig2}}
\end{figure}

\begin{figure}
\epsscale{.45}
\plotone{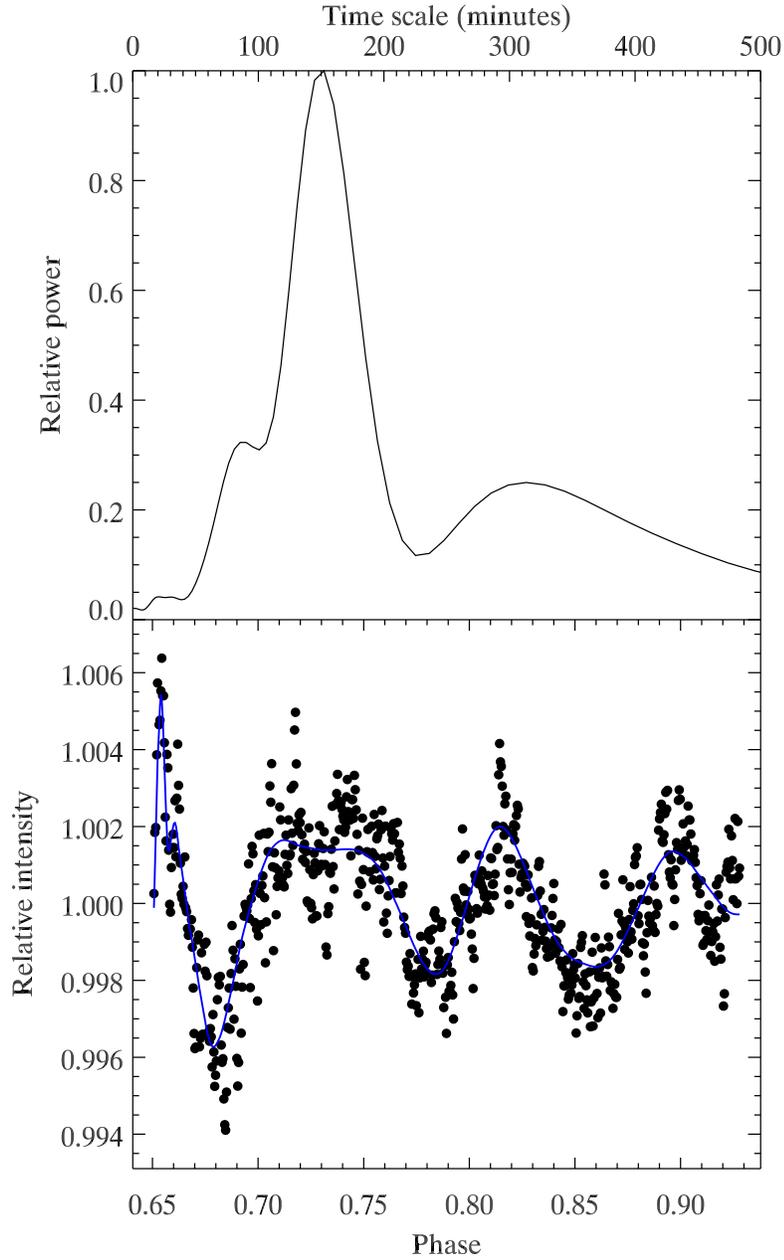}
\vspace{0.7in}
\caption{{\it Bottom panel:} Observations of WASP-33 in the J-band on
14~October, 2011, when no transit or eclipse occurred (near phase
0.8). These data have been corrected for instrument instabilities and
also corrected using the comparison stars.  Note the oscillatory
behavior of the star, with an amplitude of about $0.16\%$. The blue
line is a de-noised version using wavelets; it only incorporates the
first few wavelet coefficients so as to guide the eye to the general
variations.  {\it Top panel:} Wavelet power spectrum of the data
from the bottom panel.  The oscillatory power peaks near 146 minutes.
\label{fig3}}
\end{figure}

\clearpage

\begin{figure}
\epsscale{.45}
\plotone{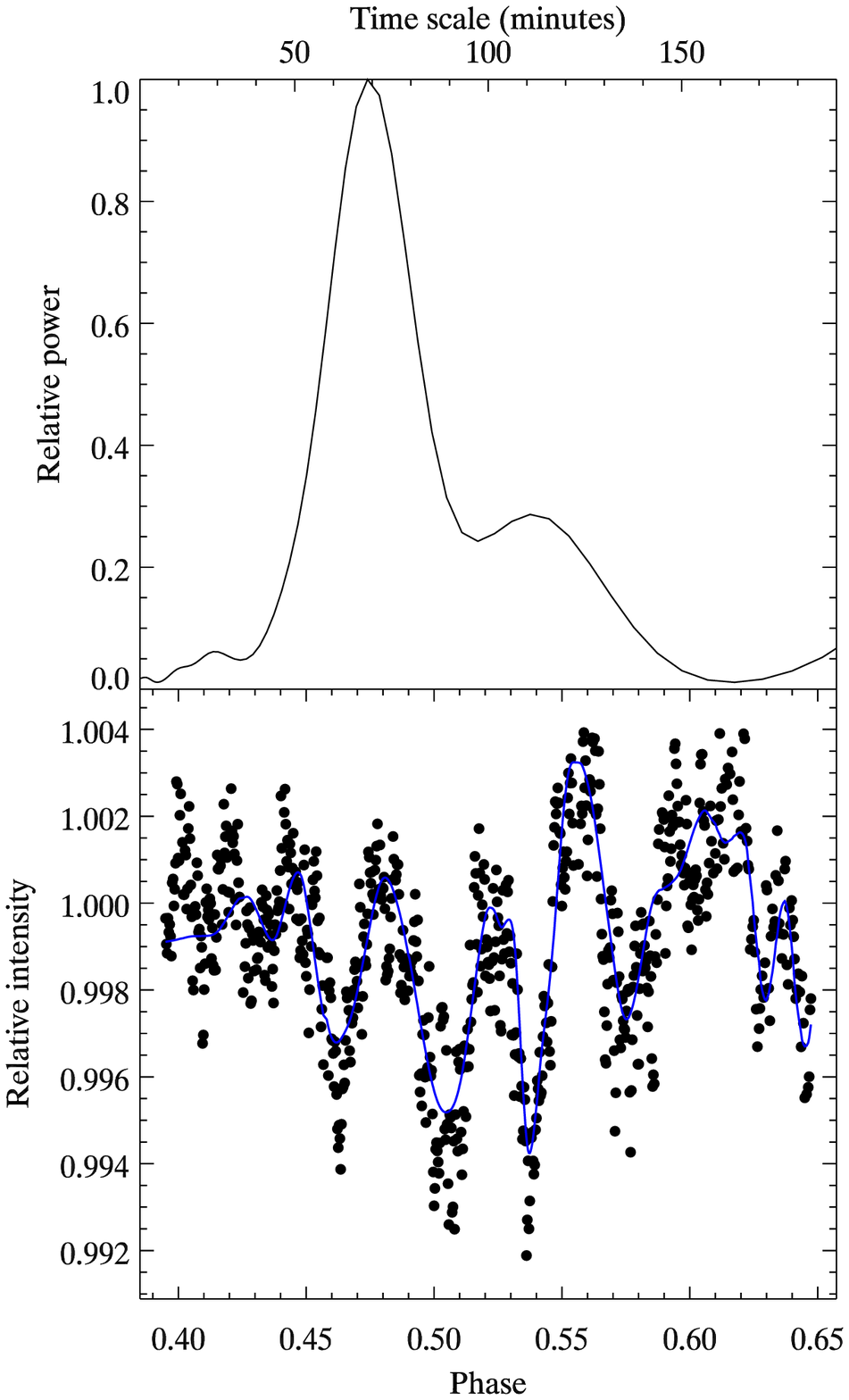}
\vspace{0.7in}
\caption{{\it Bottom panel:} Observations of WASP-33 in the $K_s$-band
on 10~October, 2011, spanning a secondary eclipse. These data have
been corrected for instrumental instabilities and also corrected using
the comparison stars. The blue line is a de-noised version using
wavelets; it only incorporates the first few wavelet coefficients so
as to guide the eye to the general variations. Note the very prominent
stellar oscillation with period near 71 minutes.  {\it Top panel:}
Wavelet power spectrum of the data from the bottom panel, showing the
oscillation peak power at 71 minutes.
\label{fig4}}
\end{figure}

\clearpage

\begin{figure}
\epsscale{.45}
\plotone{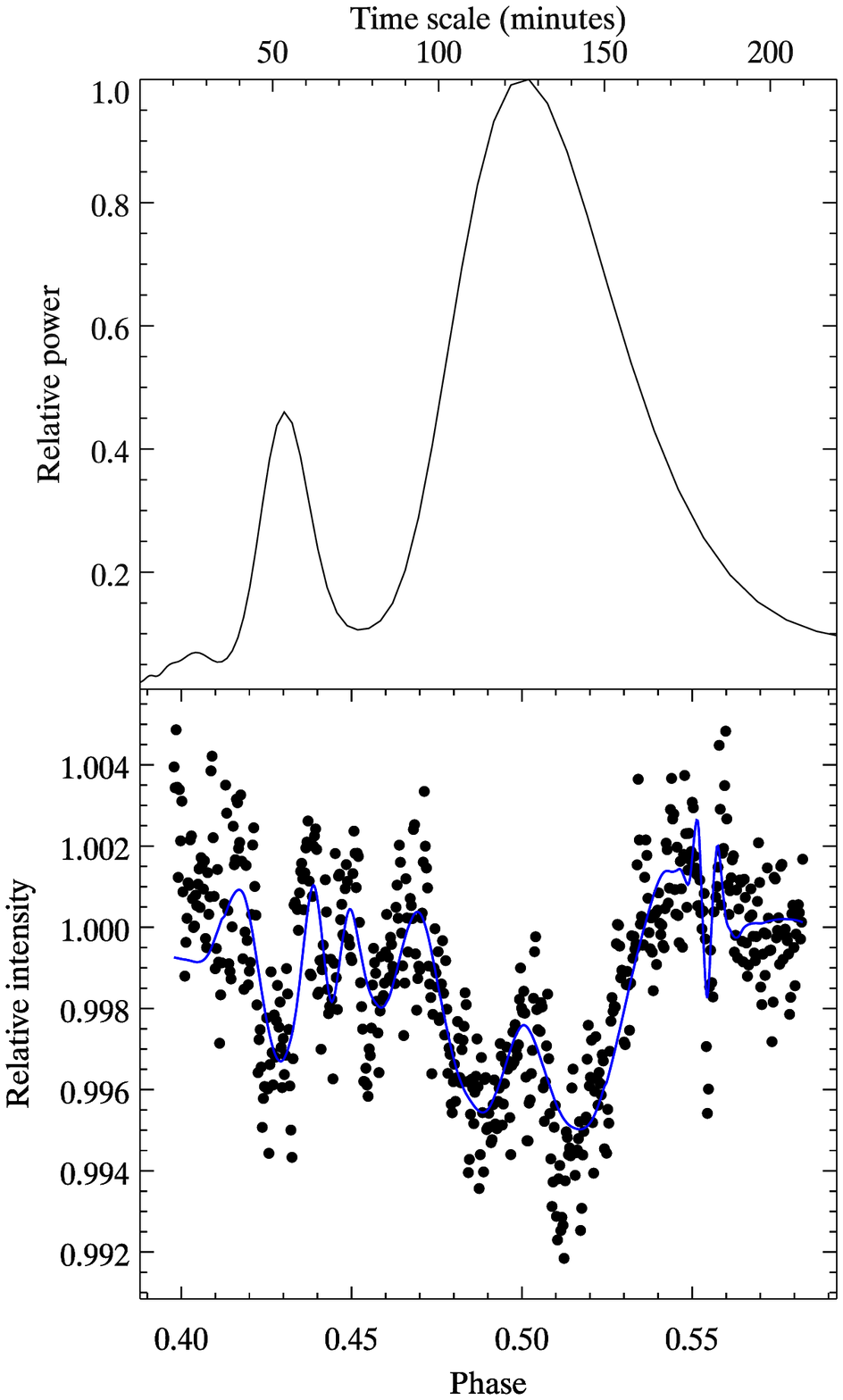}
\vspace{0.7in}
\caption{{\it Bottom panel:} Observations of WASP-33 in the $K_s$-band
on 16~October, 2011, spanning a secondary eclipse. These data have
been corrected for instrumental instabilities and also corrected using
the comparison stars. The blue line is a de-noised version using
wavelets; it only incorporates the first few wavelet coefficients so
as to guide the eye to the general variations. {\it Top panel:}
Wavelet power spectrum of the data from the bottom panel, showing
oscillatory power peaks at 54 and 126 minutes.
\label{fig5}}
\end{figure}

\clearpage

\begin{figure}
\epsscale{.50}
\plotone{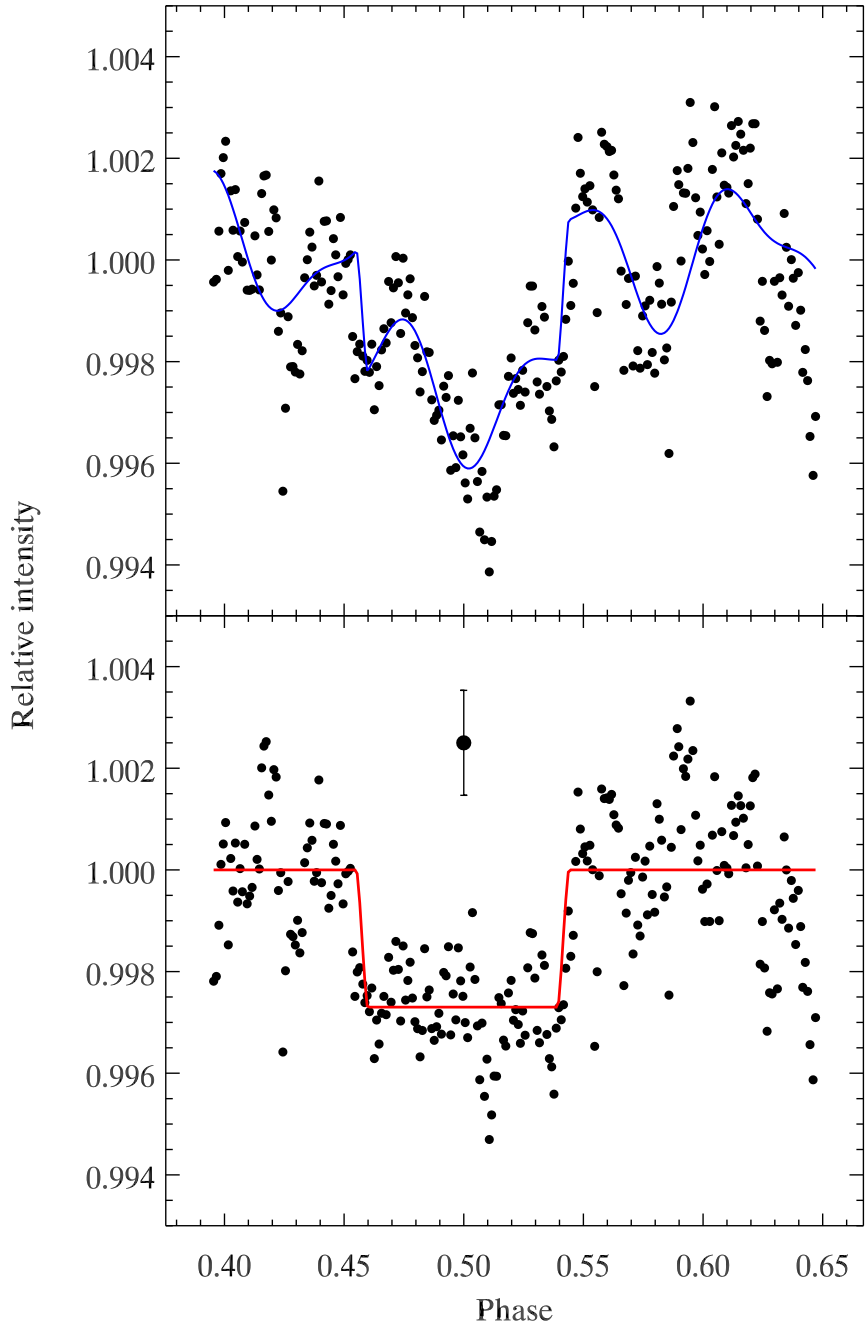}
\vspace{0.7in}
\caption{{\it Top panel:} Eclipse of WASP-33b based on combining the
$K_s$-band observations from 10 and 16 October, and binning to a phase
resolution of 0.001.  The blue line shows the result of fitting to the
eclipse, and the sum of two oscillation modes, via Markov chains. {\it
Bottom panel:} Data from the top panel with the oscillatory portion of
the fit removed and compared with the best-fit eclipse (red line). The
scatter per binned point is about 0.0012, indicated by the inset point
with error bars.  The best-fit eclipse depth is $0.0027\pm0.0002$, but
comparison of the two individual nights (not illustrated) indicates a
greater error in the depth ($\pm0.0004$, see Table~3).
\label{fig6}}
\end{figure}

\clearpage

\begin{figure}
\epsscale{.45}
\plotone{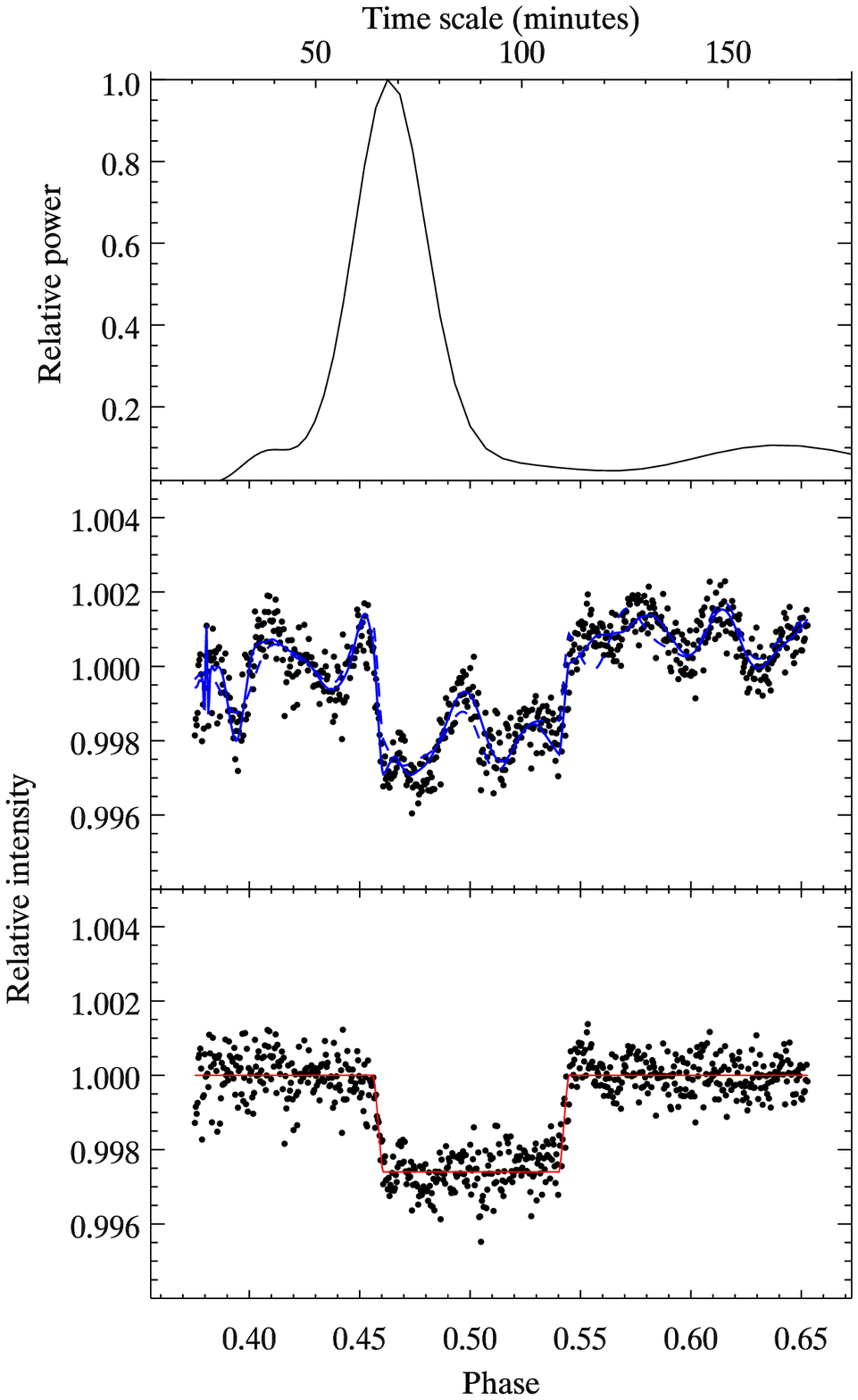}
\vspace{0.7in}
\caption{{\it Middle panel:} Spitzer observations of WASP-33 at
  3.6\,$\mu$m, spanning a secondary eclipse. The overplotted blue line
  is the best-fit solution from our MCMC analysis, including structure
  defined by our wavelet analysis (see text).  The dashed blue line
  omits the wavelet-defined structure and uses only the pure
  oscillatory portion, plus eclipse. {\it Bottom panel:} Data from the
  middle panel with the oscillatory plus wavelet portion removed,
  showing the best-fit eclipse curve (red line). {\it Top panel:}
  Wavelet power spectrum of the data points from the middle panel.  The peak
  near 68 minutes is due to oscillations of the star.
\label{fig7}}
\end{figure}

\clearpage

\begin{figure}
\epsscale{.45}
\plotone{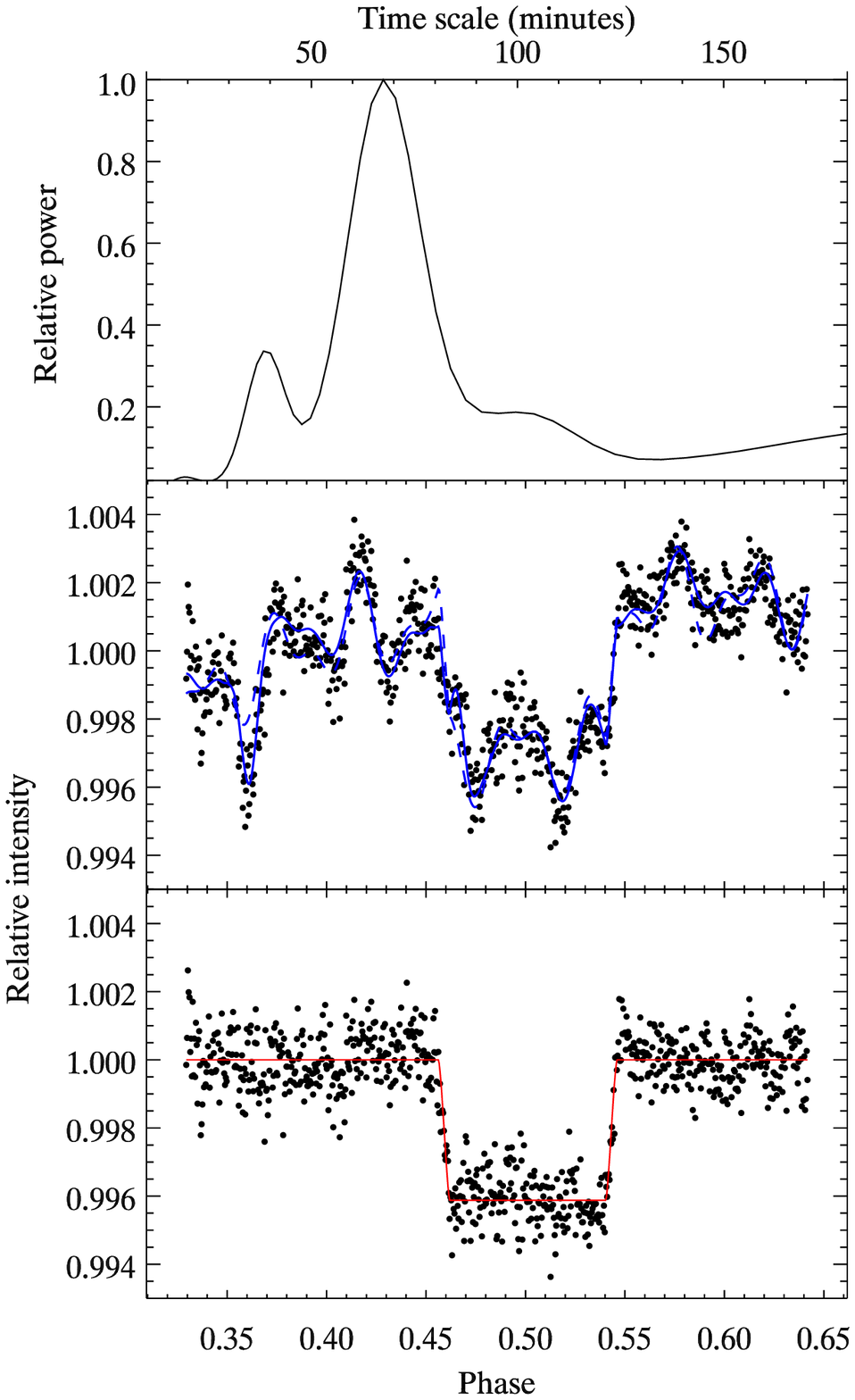}
\vspace{0.7in}
\caption{{\it Middle panel:} Spitzer observations of WASP-33 at
  4.5\,$\mu$m, spanning a secondary eclipse. The overplotted blue line
  is the best-fit solution from our MCMC analysis, including structure
  defined by our wavelet analysis (see text).  The dashed blue line
  omits the wavelet-defined structure and uses only the pure
  oscillatory portion, plus eclipse. {\it Bottom panel:} Data from the
  middle panel with the oscillatory plus wavelet portion removed,
  showing the best-fit eclipse curve (red line). {\it Top panel:}
  Wavelet power spectrum of the data from the middle panel; the main
  peak near 68 minutes is due to oscillations of the star, and a
  secondary peak near 39 minutes is due to a pointing oscillation in
  the telescope.
\label{fig8}}
\end{figure}

\clearpage

\begin{figure}
\epsscale{.45}
\plotone{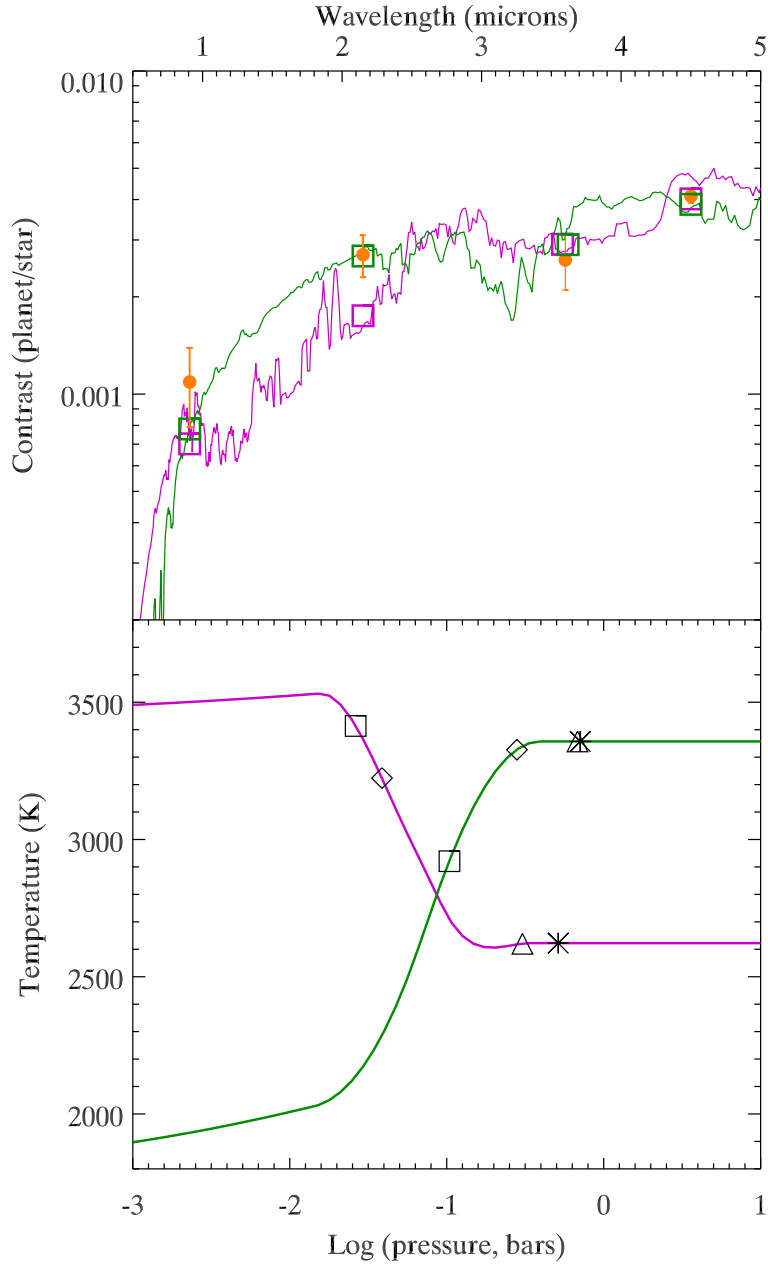}
\vspace{0.7in}
\caption{{\it Top panel:} Comparison of our measured eclipse depths
  (red-orange points at 2.15, 3.6, and 4.5\,$\mu$m, and including
  \citealp{smith} at 0.9\,$\mu$m) with two models for the atmosphere
  of WASP-33b: the green line is a carbon-rich non-inverted model, and
  the violet line is a solar composition model with an inverted
  temperature structure. The squares are the values expected by
  integrating the planetary fluxes over the bandpasses.  {\it Bottom
  panel:} Pressure-temperature profiles for the models whose emergent
  spectra are shown in the top panel. The peaks of the contribution
  functions for each bandpass are plotted as points on the
  pressure-temperature profiles.  Squares are 0.9\,$\mu$m, asterisks
  are 2.1\,$\mu$m, triangles are 3.6\,$\mu$m, and diamonds are 4.5\,$\mu$m. 
\label{fig9}}
\end{figure}

\clearpage

\begin{table}
\begin{center}
\caption{Comparison stars used for J- and $K_s$-band photometry \label{tbl-1}}
\begin{tabular}{lcll}
\tableline
\tableline
  i  & 2MASS Designation & J-mag & K-band mag \\
\tableline
 1  & 02263012+3724227 & 7.416 &  6.811  \\
 2  & 02273482+3728084 & 7.886 &  7.380  \\ 
 3  & 02263566+3735479 & 8.541 &  8.304  \\ 
 4  & 02262831+3732264 & 9.376 &  8.579  \\
 5  & 02265167+3732133 & 9.958 &  9.937   \\ 
 6  & 02271232+3728286 & 9.808 &  9.536   \\

\tableline
\end{tabular}
\end{center}
Note: All stars were used for J-band photometry, but for $K_s$-band we
omitted star 1 on 10 October because it was too bright, and we omitted
star 5 on 16 October because it was not sufficiently above the higher
thermal background on that night.  Also, star 6 was too faint for $K_s$-band
on both 10 and 16 October. For reference, WASP-33 has J-mag=7.581 and
K-mag=7.468.
\end{table}

\clearpage

\begin{sidewaystable}
\begin{center}
\caption{WASP-33 stellar infrared oscillation amplitudes and time scales \label{tbl-2}}
\begin{tabular}{lllrl}
\tableline
\tableline
 UT Date (2011) &  Wavelength ($\mu$m)  & Oscillation amplitude & Time scale(s) & Observatory  \\
\tableline
 26 March   & 3.6   & 0.0013 & 68 min           & Warm Spitzer; Figure~8  \\
 30 March   & 4.5   & 0.0013 & 68 min        & Warm Spitzer; Figure~7   \\ 
 10 October & 2.15  & 0.0023 & 71 min            & Kitt Peak 2-meter; Figure~4  \\ 
 14 October & 1.25  & 0.0016 & 146 min            & Kitt Peak 2-meter; Figure~3  \\
 16 October & 2.15  & 0.0021 & 54, 126 min        & Kitt Peak 2-meter; Figure~5   \\ 
\tableline
\end{tabular}
\end{center}
Note: \citet{smith} found oscillation periods between 42 and 77
minutes, and \citet{herrero} found a dominant period near 68 minutes.
\end{sidewaystable}

\clearpage

\begin{sidewaystable}
\begin{center}
\caption{WASP-33b secondary eclipse depths, time of central eclipse, and orbital phase  \label{tbl-3}}
\begin{tabular}{lllll}
\tableline
\tableline
 Wavelength ($\mu$m)  & Eclipse Depth & Time as BJD(TDB) & Phase & Comment  \\
\tableline
 2.15 & $0.0027\pm0.0004$ & $2455844.8156\pm0.0040$ & $0.4995\pm0.0035$  & average of 10 and 16 October; Figure~6   \\
 3.6  & $0.0026\pm0.0005$ & $2455647.1978\pm0.0001$ & $0.50041\pm0.00008$  & Warm Spitzer; Figure~7  \\
 4.5  & $0.0041\pm0.0002$ & $2455650.8584\pm0.0005$ & $0.5012\pm0.0004$  & Warm Spitzer; Figure~8  \\
\tableline
\end{tabular}
\end{center}
Note: For the ground-based eclipse at 2.15\,$\mu$m the barycentric
time of central eclipse is based on the average phase for both 10 and
16~October, then converted to the central BJD value for the 10 October
eclipse.
\end{sidewaystable}

%% If the table is more than one page long, the width of the table can vary
%% from page to page when the default \tablewidth is used, as below.  The
%% individual table widths for each page will be written to the log file; a
%% maximum tablewidth for the table can be computed from these values.
%% The \tablewidth argument can then be reset and the file reprocessed, so
%% that the table is of uniform width throughout. Try getting the widths
%% from the log file and changing the \tablewidth parameter to see how
%% adjusting this value affects table formatting.

%% The \dataset{} macro has also been applied to a few of the objects to
%% show how many observations can be tagged in a table.

\clearpage

%% Tables may also be prepared as separate files. See the accompanying
%% sample file table.tex for an example of an external table file.
%% To include an external file in your main document, use the \input
%% command. Uncomment the line below to include table.tex in this
%% sample file. (Note that you will need to comment out the \documentclass,
%% \begin{document}, and \end{document} commands from table.tex if you want
%% to include it in this document.)


%% The following command ends your manuscript. LaTeX will ignore any text


\begin{thebibliography}{}

\bibitem[Agol et al.(2010)] {agol} Agol,~E., Cowan,~N.~B.,
   Knutson,~H.~A., Deming,~D., Steffen,~J.~H., Henry,~G.~W., \&
   Charbonneau,~D., 2010, ApJ, 721, 1861.

\bibitem[Ballard et al.(2010)] {ballard} Ballard,~S., Charbonneau,~D., Deming,~D., Knutson,~H.~A.,
    Christiansen,~J.~L., Holman,~M.~J., Fabrycky,~D., Seager,~S., \& A'Hearn,~M.~F., 2010,
    PASP, 122, 1341.

\bibitem[Balona \& Dziembowski(2011)] {balona} Balona,~L.~A., \&
    Dziembowski,~W.~A., 2011, MNRAS, 417, 591.

\bibitem[Bean et al.(2011)] {bean} Bean,~J.~L., \& 9 co-authors, 2011, ApJ, 743, id.92.

\bibitem[Beerer et al.(2011)] {beerer} Beerer,~I.~M., \& 12 co-authors, 2011, ApJ, 727, id.23.

\bibitem[Berta et al.(2012)] {berta12} Berta,~Z.~K., and 9 co-authors, 2012, ApJ, 747, id.35.

\bibitem[Borysow et al.(1997)] {borysow97} Borysow,~A., Jorgensen,~U.~G., \& Zheng,~C., 1997, 
    A\&A, 324, 185.

\bibitem[Borysow(2002)] {borysow02} Borysow,~A., 2002, A\&A, 390, 779.

\bibitem[Breger \& Pamyatnykh(2006)] {breger06} Breger,~M., \&  Pamyatnykh,~A.~A., 2006, MNRAS, 368, 571.

\bibitem[Breger et al.(2009)] {breger09} Breger,~M., Lenz,~P., \& Pamyatnykh,~A.~A., 2009, MNRAS, 396, 291.

\bibitem[Burrows et al.(2007)] {burrows07} Burrows,~A., Hubeny,~I., Budaj,~J., Knutson,~H.~A., 
     \& Charbonneau,~D., 2007, ApJ, 668, L171.

\bibitem[Campo et al.(2011)] {campo} Campo,~C.~J., \& 19 co-authors, ApJ, 727, id.125.

\bibitem[Carey et al.(2010)] {carey} Carey,~S., \& 19 co-authors, 2010, SPIE, 7731, 77310N-77310N-15.

\bibitem[Castelli \& Kurucz(2004)] {kurucz} Castelli,~F., \& Kurucz,~R.~L., 2004, Proc. IAU Symposium 210, 
  eds. N. Piskunov et al., poster A20.

\bibitem[Charbonneau et al.(2008)] {charb08} Charbonneau,~D., Knutson,~H.~A., Barman,~T., Allen,~L.~E., Mayor,~M.,
   Megeath,~S.~T., Queloz,~D., \& Udry,~S., 2008, ApJ, 686, 1341.

\bibitem[Collier-Cameron et al.(2010)] {collier-cameron} Collier-Cameron,~A., \& 18 co-authors, 2011, 
   MNRAS, 407, 507.

\bibitem[Cowan \& Agol(2011)] {cowan} Cowan,~N.~B., \& Agol,~E., 2011, ApJ, 729, id.54.

\bibitem[Cowan et al.(2012)] {cowan12} Cowan,~N.~B., Machalek,~P., Croll,~B., Shekhtman,~L.~M., Burrows,~A., 
     Deming,~D., Greene,~T., \& Hora,~J.~L., 2012, ApJ, 747, id.82.

\bibitem[Croll et al.(2011)] {croll} Croll,~B., Lafreniere,~D., Albert,~L., Jayawardhana,~R., Fortney,~J.~J., 
    \& Murray,~N., AJ, 141, id.30.

\bibitem[Crossfield et al.(2012)] {crossfield} Crossfield,~I., Hansen,~B.~M.~S., \& Barman,~T., 2012, 
   ApJ, 746, id.46. 

\bibitem[Deming et al.(2005)] {deming05} Deming,~D., Seager,~S., Richardson,~L.~J., \& Harrington,~J., 2005,
     Nature, 434, 740.

\bibitem[Deming et al.(2011a)] {deming11a} Deming,~D., \& 11 co-authors, 2011, ApJ, 726, id.95.

\bibitem[Deming et al.(2011b)] {deming11b} Deming,~D., Sada,~P.~V., Jackson,~B., Peterson,~S.~W., Agol~,E., 
    Knutson,~H.~A., Jennings,~D.~E., Haase,~F., \& Bays,~K., 2011, ApJ, 740, id.33.

\bibitem[Demory et al.(2011)] {demory11} Demory,~B.-O., \& 14 co-authors, 2011, A\&A, 533, id.A114.

\bibitem[Desert et al.(2011)] {desert11} Desert,~J.-M., \& 14 co-authors, 2011, ApJS, 197, id.11.

\bibitem[Ford(2005)] {ford} Ford,~E.~B., 2005, AJ, 129, 1706.

\bibitem[Fortney et al.(2008)] {fortney} Fortney,~J.~J., Lodders,~K., Marley,~M.~S., \& Freedman,~R.~S.,
   2008, ApJ, 678, 1419.

\bibitem[Freedman et al.(2008)] {freedman08} Freedman,~R.~S., Marley,~M.~S., \& Lodders,~K., 2008,
    ApJ(Suppl), 174, 504.

\bibitem[Freedman(2009)] {freedman09} Freedman,~R.~S., 2009, personal communication.

\bibitem[Guillot et al.(1996)] {guillot}  Guillot,~T., Burrows,~A., Hubbard,~W.~B., Lunine,~J.~I., \&
    Saumon,~D., 1996, ApJ, 459, L35.

\bibitem[Harrington et al.(2007)] {harrington} Harrington,~J., Luszcz,~S., Seager,~S., Deming,~D., \& Richardson,~L.~J.,
   2007, Nature, 447, 691.

\bibitem[Hebrard et al.(2010)] {hebrard} Hebrard,~G., \& 26 co-authors, 2010, A\&A, 516, id.A95.

\bibitem[Herrero et al.(2011)] {herrero} Herrero,~E., Morales,~J.~C., Ribas,~I., \& Naves,~R., 2011, A\&A, 526, L10.

\bibitem[Kalkofen \& Ulmschneider(1977)] {kalkofen} Kalkofen,~W., \& Ulmschneider,~P., 1977, A\&A, 57, 193.

\bibitem[Karkoschka \& Tomasko(2010)] {karkoschka10} Karkoschka,~E., \& Tomasko,~M.~G., 2010, Icarus, 205, 674.

\bibitem[Karkoschka(2011)] {karkoschka11} Karkoschka,~E., 2011, personal communication.

\bibitem[Knutson et al.(2008)] {knutson08}  Knutson,~H.~A., Charbonneau,~D., Allen,~L.~E., Burrows,~A., \&
    Megeath,~S.~T., 2008, ApJ, 673, 526.

\bibitem[Knutson et al.(2009)] {knutson09}  Knutson,~H.~A., Charbonneau,~D., Burrows,~A., O'Donovan,~F.~T., \&
    Mandushev,~G., 2009, ApJ, 691, 866.

\bibitem[Knutson et al.(2010)] {knutson10}  Knutson,~H.~A., Howard,~A.~W., \& Isaacson,~H., 2010, ApJ, 720, 1569.

%%  \bibitem[Kurucz(1979)] {kurucz} Kurucz,~R.~L., 1979, ApJ(Suppl.), 40, 1.

%% \bibitem[Landini et al.(1985)] {landini} Landini,~M., Monsignori-Fossi,~B.~C., Paresce,~F., \& Stein,~R.~A.,
%%     1985, ApJ, 289, 709.

\bibitem[Madhusudhan \& Seager(2009)] {madhu09}  Madhusudhan,~N., \& Seager,~S., 2009, ApJ, 707, 24.

\bibitem[Madhusudhan \& Seager(2010)] {madhu10} Madhusudhan,~N., \& Seager,~S., 2010, ApJ, 725, 261.

\bibitem[Madhusudhan et al.(2011a)] {madhu11a} Madhusudhan,~N., and 18 co-authors, 2011, Nature, 469, 64.

\bibitem[Madhusudhan et al.(2011b)] {madhu11b} Madhusudhan,~N., Mousis,~O., Johnson,~T.~V., \& Lunine,~J.~I., 2011, ApJ, 743, id.191.

\bibitem[Mandel \& Agol(2002)] {mandel} Mandel,~K., \& Agol,~E., 2002, ApJ, 580, L171.

\bibitem[Marmolino \& Severino(1991)] {marmolino} Marmolino,~C. \& Severino,~G., 1991, A\&A, 242, 271.

%%  \bibitem[Medupe \& Kurtz(1998)] {medupe} Medupe,~R., \& Kurtz,~D.~W., 1998, MNRAS, 299, 371.

\bibitem[Menzel \& Pekeris(1935)] {menzel} Menzel,~D.~H., \& Pekeris,~C.~L., 1935, MNRAS, 96, 77.

\bibitem[Rothman et al.(2005)] {rothman05} Rothman,~L.~S., et al., 2005, JQSRT, 96, 139.

\bibitem[Sada et al.(2010)] {sada} Sada,~P.~V., \& 8 co-authors, 2010, ApJ, 720, L215.

\bibitem[Seager \& Deming(2010)] {seager-deming}  Seager,~S., \& Deming,~D., 2010, ARAA, 48, 631. 

\bibitem[Showman et al.(2008)] {showman} Showman,~A.~P., Cooper,~C.~S., Fortney,~J.~J., \& Marley,~M.~S., 2008,
    ApJ, 682, 559.

\bibitem[Smith et al.(2011)] {smith} Smith,~A.~M.~S., Anderson,~D.~R., Skillen,~I., Collier-Cameron,~A., \& 
   Smalley,~B., 2011, MNRAS, 416, 2096.

\bibitem[Spiegel et al.(2009)] {spiegel} Spiegel,~D.~S., Silverio,~K., \& Burrows,~A., 2009, ApJ, 699, 1487. 

\bibitem[Todorov et al.(2012)] {todorov} Todorov,~K.~O., \& 13 co-authors, 2012, ApJ, 746, id.111.

\bibitem[Torrence \& Compo(1998)] {torrence} Torrence,~C. \& Compo,~G.~P., Bull. Amer. Meteor. Soc., 79, 61.

\bibitem[Zahnle et al.(2009)] {zahnle} Zahnle,~K., Marley,~M.~S., Freedman,~R.~S., Lodders,~K., \&
  Fortney,~J.~J., 2009, ApJ, 701, L20. 

\bibitem[Zhao et al.(2012)] {zhao} Zhao,~M., Monnier,~J.~D., Swain,~M.~R., Barman,~T., \& Hinkley,~S., ApJ, 744, id.122.

\end{thebibliography}
\end{document}